\documentclass[pra,twocolumn,aps,nobibnotes]{revtex4}
\usepackage{graphicx}%
\usepackage{dcolumn}
\usepackage{amsmath}   
\usepackage{bm}
\usepackage{longtable}
\usepackage{pstricks,pst-node,pst-coil}

\begin{document}
\title{Activation entropy of electron transfer reactions}
\author{Anatoli A.\ Milischuk, Dmitry V.\ Matyushov,}
\email[E-mail:]{dmitrym@asu.edu.}
\affiliation{ 
Department of Chemistry
and Biochemistry and the Center for the Early Events in Photosynthesis, 
Arizona State University, PO Box 871604, Tempe, 
AZ 85287-1604}
\author{and Marshall D.\ Newton}\email[E-mail:]{newton@bnl.gov}
\affiliation{Brookhaven National Laboratory, Chemistry Department, Box 5000,
             Upton, New York 11973-5000}
\date{\today}
\begin{abstract}
  We report microscopic calculations of free energies and entropies for
  intramolecular electron transfer reactions. The calculation
  algorithm combines the atomistic geometry and charge distribution of
  a molecular solute obtained from quantum calculations with the
  microscopic polarization response of a polar solvent expressed in
  terms of its polarization structure factors.  The procedure is
  tested on a donor-acceptor complex in which ruthenium donor and
  cobalt acceptor sites are linked by a four-proline polypeptide.  The
  reorganization energies and reaction energy gaps are calculated as a function
  of temperature by using structure factors obtained from our
  analytical procedure and from computer simulations. Good agreement
  between two procedures and with direct computer simulations of the
  reorganization energy is achieved. The microscopic algorithm is
  compared to the dielectric continuum calculations. We found that the
  strong dependence of the reorganization energy on the solvent
  refractive index predicted by continuum models is not supported by
  the microscopic theory. Also, the reorganization and overall
  solvation entropies are substantially larger in the microscopic
  theory compared to continuum models.
\end{abstract}
\maketitle

\section{Introduction}\label{sec:1}
Beginning with work of Marcus on electron transfer (ET) between ions
dissolved in polar solvents \cite{Marcus:93}, the understanding of the
dynamics and thermodynamics of the nuclear polarization coupled to the
transferred electron has been viewed as a key component of ET
theories.  The concept of polarization fluctuations as a major
mechanism driving ET has been extended over the several decades of
research from simple molecular solvents to a diversity of
condensed-phase media of varying complexity.  A significant part of
the present experimental and theoretical effort is directed toward the
understanding of ET in biology, where this process is a key component
of energy transport chains
\cite{MarcusSutin,Winkler:92,Mclendon:92,Warshel:02}.  Biological
systems pose a major challenge to theoretical and computational
chemistry from at least two viewpoints.  First, the solvent, including
bulk and bound water \cite{Gregory:95}, membranes, and parts of the
polar and polarizable matrix of the biopolymer, is highly anisotropic
and heterogeneous.  Second, the geometry of what can be
separated as a solute is often very complex, including concave regions
of molecular scale occupied by the solvent and regions of the
biopolymer with a significant mobility of polar and ionizable
residues.

Dielectric continuum models accommodate the complex solute shape by
numerical algorithms solving the Poisson equation with the boundary
conditions defined by a dielectric cavity \cite{delphi}. The
heterogeneous nature of the solvent in the vicinity of a redox site
can in principle be included by assigning different dielectric
constants to its heterogeneous parts \cite{Siriwong:03}. Two
fundamental problems inevitably arise in this algorithm. The first has
been well recognized over the years of its application and is related
to the ambiguity of defining the dielectric cavity for molecular
solutes.  This problem is often resolved by proper parameterization of
the radii of atomic and molecular groups of the solute. The second
problem is much less studied. It is related to the fact that
collective polarization fluctuations of molecular dielectrics possess
a finite correlation length which may be comparable to the length of
concave regions of the solute or some other characteristic dimensions
significant for solvation thermodynamics. The definition of the
dielectric constant for polar regions of molecular length is very
ambiguous and, in addition, once the dielectric constant is defined,
it is not clear if the dielectric response can fully develop on the
molecular length scale.

In addition to the problems in implementing the continuum formalism for
molecular solutes there are some fundamental limitations of the
continuum approximation itself that may limit its applicability to
solvation and electron transfer thermodynamics. On the basic level,
the definition of the molecular cavity should be re-done for each
particular thermodynamic state of the solvent
\cite{Roux:90,LyndenBell:99} and/or electronic state of the solute
\cite{Rick:94}. This precludes the use of continuum theories with a
given cavity parametrization to describe derivatives of the solvation
free energy, e.g. entropy and volume of solvation \cite{DMjpcb:99}. In
addition, the calculation of the free energy of ET activation requires
a proper separation of nuclear solvation from the overall solvent
response. This problem, actively studied by formal theories in the
past \cite{Lee:88,Kuznetsov:92,Gehlen:92,Zhu:95}, has been recently addressed by computer
simulations \cite{Bader:96,Ando:01,DMjpca:04}.  Computer simulations
have indicated that continuum recipes for the separation of nuclear
and electronic polarization are unreliable, resulting in too strong a
dependence of the solvent reorganization (free) energy on solvent refractive
index. All these limitations call for an extension of traditional
approaches to solvation and ET thermodynamics that would include
microscopic length-scales of solvent polarization.

Microscopic theories of solvation are not yet sufficiently developed
to compete efficiently with continuum models in application to
solvation of biopolymers. Computer simulations provide a very detailed
picture of the local solvation structure, but their application to
solvation of large solutes requires very lengthy computations and
often includes approximations that are hard to control.  In
particular, the dielectric response is very slowly converging in
simulations and is potentially affected by approximations made to
describe the long-range electrostatic forces.  Several simulation
protocols in which polarization response is (partially) integrated out
by analytical techniques have been proposed
\cite{Marchi:01,Leontyev:03}.  Integral equation theories have been
successfully applied to small solutes \cite{Raineri:99}, but examples
of their application to solvation and reactivity of large solutes are
just a few \cite{Beglov:96}. The formulation of the solvation problem
in terms of molecular response functions holds significant promise,
as it combines the molecular length scale of the polarization response
with a possibility to accommodate an arbitrary shape of the solute
\cite{Kornyshev:85,Kornyshev:86,Fried:90,Bagchi:91,Chandler:93,DMcp:93,Song:96,Kornyshev:96,Song:98,Lang:99,Ramirez:02}.
A recent re-formulation of the Gaussian model \cite{Chandler:93} for
solvation in polar solvents \cite{DMjcp1:04,DMjcp2:04} shows a good
agreement with simulations of model systems and an ability to conform
with experiment when applied to ET in biomolecules \cite{DMjpcb1:03}
and charge-transfer complexes \cite{DMjcp4:05} and to solvation
dynamics \cite{DMjcp1:05}.  Testing the algorithm, referred to as the
non-local response function theory (NRFT), on model systems for which
both computer simulations and experiment exist is critical for future
applications to more complex systems. This is the aim of the present
contribution.

\begin{figure}[htbp]
  \centering  \includegraphics*[width=6cm]{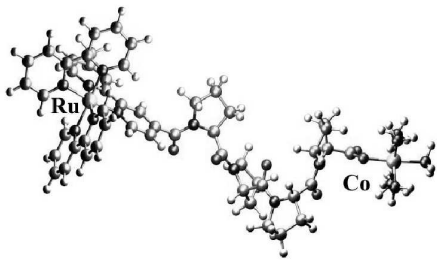}
  \caption{Diagram of the polypeptide donor-spacer-acceptor (DSA) complex
   referred to as complex \textbf{1} in the text.}
  \label{fig:1poly}
\end{figure}

Testing microscopic solvation theories requires comparison to computer
simulations on model, yet realistic, systems.  The current
experimental database does not provide sufficient accuracy to test
various approximations entering theoretical algorithms. On the
other hand, computer experiment offers essentially exact (within the
accuracy of simulation protocols) integration of the same Hamiltonian
as the one used in the analytical theory.  Therefore, the present
calculations of the ET thermodynamics are compared to recent very
extensive Molecular Dynamics (MD) simulations \cite{Ungar:99} of a
donor-spacer-acceptor (DSA) complex consisting of transition-metal
donor (D) and acceptor (A) sites linked by a polyproline peptide
spacer (S) (Fig.\ \ref{fig:1poly}):
\[
\mathrm{(bpy)_2Ru^{2+}(bpy')-(pro)_4-O^-Co^{3+}(NH_3)_5 },
\] 
where in the donor bpy=2,2$'$-bipyridine and
bpy$'$=4$'$-methyl-2,2$'$-bipyridyl.  The spacer is a polyproline
chain whose first member (the N-terminus) is connected to the bpy$'$
carbonyl, and whose fourth member is terminated by a carboxylate
moiety bound to the $\mathrm{-Co^{3+}(NH_3)_5}$ acceptor. This system,
modeling ET in redox proteins, is a representative member of a
homologous series of DSA complexes for which ET rates as a function of
temperature have been reported \cite{Ogawa:93}. This complex will be
referred to as complex \textbf{1} in the text.

The analytical NRFT model is shown to agree exceptionally well with MD
simulations for complex \textbf{1} (Figure \ref{fig:1poly}) in TIP3P
water.  In order to provide a rigorous comparison between simulations
and analytical theory, the set of solute charges employed in the simulations was
also used in the analytical calculations. In addition, the
polarization structure factors of TIP3P water were obtained from
separate MD simulations to be used as input in the analytical theory.
Once the accuracy and robustness of the analytical procedures are
tested on MD simulations, the next step is to see if the model is
capable of reflecting the behavior of real systems. To this end, we have
developed a parameterization scheme for polarization structure factors
applicable to polarizable polar solvents. Once this is done, the
theory can be extended to calculations at varying thermodynamic
conditions of the solvent (e.g., temperature) and should generate a
set of predictions which can be tested experimentally.

We use the polypeptide DSA to focus on two problematic areas of
dielectric continuum models: dependence of the reorganization energy
on the solvent polarizability \cite{Bader:96,DMjpca:04} and the
entropy of nuclear solvation \cite{DMcp:93,DMjpcb:99}. For both areas
there is a fundamental, both quantitative and qualitative,
disagreement between microscopic models and continuum calculations.
Unfortunately, no experimental evidence on the dependence of the
solvent reorganization energy on solvent refractive index is available
in the literature. There is, on the other hand, a limited number of
experimental
\cite{Grampp:84,Liang:89,Dong:92,Derr:98,Nelsen:99,Derr:99,DMjpcb:99,Vath:00,Zhao:01,Coropceanu:03,Mertz:05}
and simulation \cite{Leontiev:05} studies on the entropy of
reorganization.  Most of the available experimental (laboratory and
simulation) evidence points to a positive reorganization entropy
(i.e., a negative slope of the reorganization energy vs temperature)
in polar solvents, in agreement with the prediction of microscopic
theory \cite{DMcp:93} and in disagreement with negative entropies from
continuum calculations \cite{Kumar:98,DMjpcb:99}.  We are aware,
however, of a few measurements performed on charged donor-acceptor
complexes indicating either zero or negative reorganization entropies
\cite{Dong:92,Coropceanu:03,Mertz:05}.  Our current calculations on
complex \textbf{1} (Fig.\ \ref{fig:1poly}) give absolute values of the
reorganization entropy much higher than continuum calculations. This
great discrepancy calls for additional tests of the theory against
experimental data, which will be a subject of future work.

\section{Golden Rule Rate Constant}
\label{sec:2}
The Golden Rule rate constant of ET is \cite{Kubo:55} 
\begin{equation}
  \label{eq:0}
  k_{\text{ET}} = \frac{2\pi V_{12}^2}{\hbar^2} \mathrm{FCWD}(0) ,
\end{equation}
where FCWD stands for the density-of-states weighted Franck-Condon 
(FC) factor
\begin{equation}
  \label{eq:0-1}
  \mathrm{FCWD}(\omega) = 
  \int \frac{dt}{2\pi} \left\langle  e^{iH_2t/ \hbar} e^{-iH_1t/ \hbar}\right\rangle_n e^{-i \omega t}.  
\end{equation}
Here, $\langle\dots\rangle_n$ is an ensemble average over the nuclear degrees of
freedom of the system (denoted by subscript ``n''), which include the
manifold of $N$ normal vibrational modes of the donor-acceptor complex
$Q=\{\mathbf{q}_1,\dots\mathbf{q}_N\}$ and the nuclear component of the
dipolar polarization of the solvent $\mathbf{P}_n$.  The ensemble
average is carried out over the configurations in equilibrium with the
initial state.  Further, $H_{i}$ ($i=1,2$) are the diagonal matrix
elements of the unperturbed system Hamiltonian $H$ taken on the
two-state electronic basis $\{\Psi_1,\Psi_2 \}$: $H_{i}=\langle\Psi_{i}|H|\Psi_{i}\rangle$
($i=1$ and $i=2$ stand for the initial and final electronic states,
respectively). The sum of $H$ and the perturbation $V$ makes the whole
system Hamiltonian, $H'=H+V$, and $V_{12} = \langle\Psi_1|V|\Psi_2\rangle$ is the
off-diagonal matrix element in the Golden Rule expression.

The system Hamiltonian of a donor-acceptor complex in a
condensed-phase solvent can be separated into the gas-phase component,
$H_g$, the solute-solvent interaction, $H_{0s}$ (``0'' stands for the
solute, ``s'' stands for the solvent), and the bath Hamiltonian,
$H_B$, describing thermal fluctuations of the solvent:
\begin{equation}
  \label{eq:1-2}
  H = H_g + H_{0s} + H_{B} .
\end{equation}
The gas-phase Hamiltonian is the sum of the kinetic energy of the
electrons, kinetic energy of the nuclei, and the full electron-nuclear
Coulomb energy.  The solute-solvent Hamiltonian for ET in dipolar
solvents is commonly given by the coupling of the operator of the
solute electric field $\mathbf{\hat E}_0$ to the dipolar polarization
of the solvent $\mathbf{P}$
\begin{equation}
  \label{eq:1-4}
     H_{0s}  =  - \mathbf{\hat E}_0*\mathbf{P} . 
\end{equation}
The bath Hamiltonian represents Gaussian statistics of the
collective mode $\mathbf{P}$ with the linear response function
$\bm{\chi}(\mathbf{r},\mathbf{r}')$
\begin{equation}
  \label{eq:1-5}
  H_B = \frac{1}{2}\mathbf{P}*\bm{\chi}^{-1}*\mathbf{P} .
\end{equation}
The asterisk between the bold capital letters denotes tensor
contraction (scalar product for vectors) and space integration over
the volume $\Omega$ occupied by the solvent
\begin{equation}
  \label{eq:1-6}
    \mathbf{E}*\mathbf{P} = \int_{\Omega} \mathbf{E}\cdot \mathbf{P} d\mathbf{r} .
\end{equation}

Assuming that the intramolecular vibrations are decoupled from solvent
nuclear modes allows one to cast the FCWD as a convolution of the
vibrational, $G_v(\omega)$, and solvent, $G_s(\omega)$, FC densities
\cite{BixonJortner:99}:
\begin{equation}
  \label{eq:1-7}
  \mathrm{FCWD}(\omega) = 
   \int_{-\infty}^{\infty} d\omega' G_v(\omega')G_s(\omega-\omega'-\Delta G/ \hbar) ,
\end{equation}
where the diabatic equilibrium free energy gap is the sum of the gas-phase component
$\Delta G_g$ and difference in solvation energies $\Delta G_s$ 
\begin{equation}
  \label{eq:1-7-1}
  \Delta G = \Delta G_g + \Delta G_s .
\end{equation}
In the absence of vibrational frequency change, the former component is 
equal to the 0-0 transition energy in the gas phase.

The FC density for each nuclear mode is given in terms of
a broadening function $g_n(t)$ \cite{Ovchinnikov:69,Mukamel:95}
\begin{equation}
  \label{eq:1-8}
  G_n(\omega') = \int_{-\infty}^{\infty} \frac{dt}{2\pi} \exp\left[i(\lambda_n/ \hbar -\omega')t - g_n(t) \right],
\end{equation}
where
\begin{equation}
  \label{eq:1-9}
  \begin{split}
  g_n(t) = & \frac{1}{\pi} \int_0^{\infty} \frac{dz}{z^2}(1-\cos z t)\chi_n''(z) \coth \frac{\hbar z}{2k_{\text{B}}T}\\
          & + \frac{i}{\pi}\int_0^{\infty}\frac{dz}{z^2}(z t - \sin z t)\chi_n''(z) .
  \end{split}
\end{equation}
In Eq.\ (\ref{eq:1-8}), $\lambda_n$ is the nuclear reorganization energy
\begin{equation}
  \label{eq:1-10}
  \lambda_n = \frac{\hbar}{\pi} \int_0^{\infty} \frac{d z}{z} \chi_n''(z)
\end{equation}
and $\chi_n''(z)$ is the imaginary part of the frequency-dependent linear
response function (spectral density) corresponding to the nuclear mode
$n$ (in general, many such modes contribute to the solvent (s) and
vibrational (v) FC densities).

For a set of vibrational normal modes with frequencies $\omega_q$ and reorganization
energies $\lambda_q$, the spectral density is \cite{Mukamel:95}
\begin{equation}
  \label{eq:1-11}
  \chi_v''(z) = \pi \sum_q S_q\omega_q^2\left[\delta(z-\omega_q) - \delta(z+\omega_q) \right],
\end{equation}
where $S_q=\lambda_q/ \hbar \omega_q $ is the Huang-Rhys factor.  When all
nuclear modes are classical, $g_n(t) = k_{\text{B}}T \lambda_n t^2/ \hbar^2$
and one reaches the classical, high temperature limit of the Marcus
theory
\begin{equation}
  \label{eq:1-12}
  \begin{split}
   \mathrm{FCWD}_1(\omega )= & \left[4\pi(\lambda_s+\lambda_v)k_{\text{B}}T\right]^{-1/2} \\
      & \exp\left[- \frac{(\Delta G+ \lambda_s +\lambda_v - \hbar \omega )^2}{4k_{\text{B}}T(\lambda_s+\lambda_v)} \right] ,
  \end{split}
\end{equation}  
where $\lambda_v=\sum_q\lambda_q$ is the total vibrational reorganization energy. 

When the solvent mode is classical and the vibrations are quantized, one
can use the small $t$ expansion in Eq.\ (\ref{eq:1-8}), valid in the
limit when $\omega_q$ is much smaller than the vertical energy gap
[$|\lambda_v - \omega'|$ in Eq.\ (\ref{eq:1-8})].  With $\hbar\omega_q / k_{\text{B}}T \gg
1$, one gets
\begin{equation}
  \label{eq:1-13}
  g_v(t) \simeq  (t^2/2\hbar)\omega_v\lambda_v,
\end{equation}
where 
\begin{equation}
  \label{eq:1-14}
  \omega_v = (\lambda_v)^{-1} \sum_q \omega_q \lambda_q 
\end{equation}
is the effective vibrational frequency. With the vibrational
broadening function in the form of Eq.\ (\ref{eq:1-13}) the FCWD becomes
\cite{Holstein:59,Hopfield:74,Siders:81,Marcus:89}
\begin{equation}
  \label{eq:1-15}
  \begin{split}
        \mathrm{FCWD}_1(\omega)= & 
        \left[\pi(4\lambda_sk_{\text{B}}T+2\hbar\omega_v\lambda_v)\right]^{-1/2} \\
      & \exp\left[-\frac{(\Delta G+ \lambda_s +\lambda_v - \hbar \omega )^2}
                        {4k_{\text{B}}T\lambda_s+2\hbar\omega_v\lambda_v} \right] .
  \end{split}
\end{equation}  
The above equation, present in some early papers on ET
\cite{Hopfield:74,Siders:81}, is not very accurate as was pointed out
by Jortner \cite{Jortner:76}. The set of equations given below, which
can be found in work by Lax \cite{Lax:52}, Davydov \cite{Davydov:53},
and Kubo and Toyozawa \cite{Kubo:55}, provides a better description of
the vibronic envelope.

When the normal mode vibrations are represented by a single effective vibration
with frequency defined by Eq.\ (\ref{eq:1-14}), the vibrational
FCWD is a weighted sum of resonant vibrational transitions
\begin{equation}
  \label{eq:1-16}
   G_v(\omega) = \sum_{m=-\infty}^{\infty} A_m \delta(\omega - m\omega_v), 
\end{equation}
where
\begin{equation}
  \label{eq:1-17}
A_m= e^{-S\coth \chi_v + m\chi_v} I_m\left( \dfrac{S}{\sinh \chi_v}\right)  ,
\end{equation}
$S=\lambda_v/\hbar\omega_v$, $\chi_v=\hbar\omega_v/2k_{\text{B}}T$, and $I_m(x)$ is the modified
Bessel function of order $m$.

The FCWD for the classical nuclear modes of the solvent is given by the
expression
\begin{equation}
  \label{eq:1-18}
  G_s(\omega - \Delta G/ \hbar  ) = \hbar \langle \delta(\Delta E(\mathbf{P}_n) - \hbar\omega)\rangle ,
\end{equation}
where 
\begin{equation}
  \label{eq:1-19}
  \Delta E(\mathbf{P}_n) = \Delta G + \lambda_s - \Delta \mathbf{E}_0*\delta\mathbf{P}_n
\end{equation}
and $\langle\dots\rangle_1$ denotes an ensemble average over the fluctuations of
the nuclear solvent polarization $\mathbf{P}_n$ coupled to the
difference in initial and final state electric fields of the
donor-acceptor complex,
$\Delta\mathbf{E}_0=\mathbf{E}_{02}-\mathbf{E}_{01}$.  In Eq.\ 
(\ref{eq:1-19}), $\lambda_s$ stands for the solvent reorganization energy
(see below), and $\delta \mathbf{P}_n$ is the fluctuation of the nuclear
polarization with respect to its equilibrium value.  With the Gaussian
Hamiltonian for polarization fluctuations [Eq.\ (\ref{eq:1-5})],
$G_s(\omega-\Delta G/ \hbar )$ is a Gaussian function leading to a total FCWD in the
form of a weighted sum of Gaussians
\begin{equation}
  \label{eq:1-20}
  \begin{split}
  \mathrm{FCWD}_i(\omega)&= \left[4\pi\lambda_sk_{\text{B}}T\right]^{-1/2}\sum_{m=-\infty}^{\infty} 
     A_m \\
  &\exp\left(-\frac{\left(\Delta G + \lambda_s + m\hbar\omega_v - \hbar \omega \right) ^2}{4\lambda_sk_{\text{B}}T}\right) .
\end{split}
\end{equation}
When the energy of vibrational excitations is much greater than 
$k_{\text{B}}T$ [$\chi_v \gg 1$ in Eq.\ (\ref{eq:1-17})] the FC envelope turns
into a sum of Gaussians with weights given by the Poisson distribution
\cite{BixonJortner:99}
\begin{equation}
  \label{eq:1-21}
  A_m = e^{-S} \frac{S^m}{m!}, \quad m>0 .
\end{equation}

\section{Solvation Thermodynamics}\label{sec:3}
Inserting a solute into a molecular solvent results in solvent
perturbation that can roughly be split into two components with
drastically different length scales. The first component is due to
repulsion of the solvent from the solute core caused by short-range,
but strong repulsive forces. This perturbation creates a local density
profile in the solvent around the solute which may or may not induce a
polarization field acting on the solute charges. The electric field of
solute charges creates yet another perturbation. The solute electric
field is sufficiently long-ranged to induce the dipolar polarization
$\mathbf{P}(\mathbf{r})$ in a quasi-macroscopic region of the solvent
around the solute.  Gradients of the solute field couple to the
higher-order (quadrupolar, etc.)  polarization, but this interaction
is more short-ranged \cite{Perng:96,Perng2:96,DMjcp2:99,DMjcp4:05}.

The dipolar polarization is caused by alignment of the permanent and
induced solvent dipoles along the solute field. This alignment occurs
on two quite different time scales: $\simeq 10^{-15}$ s for induced dipoles and
$\simeq10^{-11}-10^{-12}$ s for permanent dipoles.  Accordingly, the
polarization field splits into a fast relaxing electronic polarization
(induced dipoles, $\mathbf{P}_e$) and a much slower nuclear
polarization (permanent dipoles, $\mathbf{P}_n$) \cite{comIR}.  The
electronic solvent polarization is always in equilibrium with the
changing distribution of the electronic density in the donor-acceptor
complex. The energy conservation condition of the Golden Rule formula
is thus imposed on the energies with equilibrated electronic
polarization.  Therefore, before being used in the Golden Rule
expression, the Hamiltonian matrix should be averaged over the fast
electronic component of the dipolar polarization
\cite{Gehlen:92,DMjcp:98}. For the energy $E_i$ depending on the
instantaneous configuration of the nuclear subsystem one gets
\begin{equation}
  \label{eq:1-21-1}
  e^{-\beta E_i} = \mathrm{Tr}_{\text{el}}\left[ e^{-H_i/k_{\text{B}}T}\right],
\end{equation}
where Tr$_{\text{el}}$ denotes the statistical average over the
electronic degrees of freedom of the solvent.  Before going into the
details of separate calculations for electronic and nuclear components
of the polarization, we outline the general formalism of polarization
response to an external electric field.

\begin{figure}[htbp]
  \centering
  \includegraphics*[width=6cm]{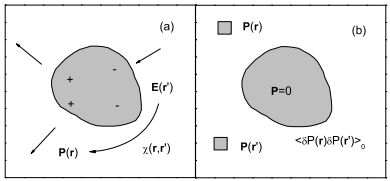}
  \caption{Two approaches to the calculation of the response function: as polarization 
    response to an external electric field perturbation (a) and as
    correlation of polarization fluctuations near the solute hard core
    from which the polarization field is excluded (b). }
  \label{fig:2poly}
\end{figure}

\subsection{Formalism}
In the linear response approximation (LRA), the solvent polarization
$\mathbf{P}(\mathbf{r})$ is a linear functional of the perturbing
electric field $\mathbf{E}_0$ (vacuum electric field of the solute for
solvation):
\begin{equation}
  \label{eq:2-1}
  \mathbf{P}(\mathbf{r}) = \bm{\chi}*\mathbf{E}_0 = \int \bm{\chi}(\mathbf{r},\mathbf{r}') \cdot
                                \mathbf{E}_0(\mathbf{r}') d\mathbf{r}' .
\end{equation}
Here, $\bm{\chi}(\mathbf{r},\mathbf{r}')$ is a two-rank tensor describing
the non-local linear response of the solvent to the solute electric
field, dot denotes tensor contraction over the common Cartesian
projections.  This function is different from dielectric
susceptibility appearing in Maxwell's equations in two respects.
First, $\bm{\chi}(\mathbf{r},\mathbf{r}')$ describes the polarization
response to the field of external charges and \textit{not} to the
total electric field $\mathbf{E}=\mathbf{E}_0+\mathbf{E}_P$ combining
the external field with the electric field $\mathbf{E}_P$ created by
the solvent polarization ($\bm{\chi}$ corresponds to $\bm{\chi}^0$ of Madden
and Kivelson \cite{Madden:84}).  Second,
$\bm{\chi}(\mathbf{r},\mathbf{r}')$ is affected by the presence of the
solute and thus $\bm{\chi}(\mathbf{r}_1,\mathbf{r}')$ is generally not
equal to $\bm{\chi}(\mathbf{r}_2,\mathbf{r}'')$ even if
$\mathbf{r}_1-\mathbf{r}'= \mathbf{r}_2 - \mathbf{r}'' $.

Equation (\ref{eq:2-1}) determines the response function in terms of
an external electrostatic perturbation and polarization induced by it  
(Fig.\ \ref{fig:2poly}a). An alternative view of the response function is
through the fluctuation-dissipation theorem which relates the response
to the correlation function of polarization fluctuations in the solute
vicinity
\begin{equation}
  \label{eq:2-2-1}
  \bm{\chi}(\mathbf{r},\mathbf{r}') = (k_{\text{B}}T)^{-1}\langle\delta\mathbf{P}(\mathbf{r})\delta \mathbf{P}(\mathbf{r}')\rangle . 
\end{equation}
An important result of the LRA is that this correlation function
does not depend on the long-range electrostatic field of the solute.
The ensemble average $\langle\dots\rangle$ in the presence of the real solute with
its charge distribution is equivalent to the ensemble average
$\langle\dots\rangle_0$ in the presence of a fictitious solute which has the
geometry of the real solute (and therefore the complete repulsion
potential) but no partial charges.  This notion provides a convenient
route to the calculations of the response function for complex
solutes. Instead of calculating the polarization in response to a
non-trivial field $\mathbf{E}_0(\mathbf{r})$, one can calculate the
correlation of polarization fluctuations in the presence of a
fictitious solute with only the hard repulsive core of the real solute
retained.  The correlation function is then calculated with the
requirement of zero polarization within the solute (Fig.\ 
\ref{fig:2poly}b)
\begin{equation}
  \label{eq:2-2-2}
  \bm{\chi}(\mathbf{r},\mathbf{r}') = 
    (k_{\text{B}}T)^{-1}\langle\delta\mathbf{P}(\mathbf{r})\delta \mathbf{P}(\mathbf{r}')\rangle_0 .
\end{equation}
This is the essence of the approach adopted in the present formalism,
making the response function solely determined by the molecular
structure inherent to the pure solvent and the short-range
perturbation produced by the repulsive core of the solute.

\begin{figure}[htbp]
  \centering
  \includegraphics*[width=6cm]{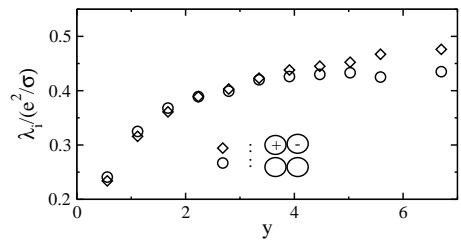}
  \caption{$\lambda_1$ for a neutral diatomic D-A (circles) and $\lambda_2$ for a polar diatomic
    D$^+$-A$^-$ (diamonds) for a donor-acceptor complex represented by
    two contact spheres with radii $R_0/ \sigma=0.9$.  Solvent is a fluid
    of dipolar hard spheres of diameter $\sigma$ and dipole moment $m$. The
    simulations, reported in Ref.\ \onlinecite{DMjcp2:04}, were
    carried out at different $m$ and at constant density $\rho\sigma^3=0.8$.
    The change in solvent polarity is reflected by the dipolar density
    $y=(4\pi/9)m^2\rho/k_{\text{B}}T$. }
  \label{fig:3poly}
\end{figure}

The applicability of the LRA to solvation of large donor-acceptor
complexes common for ET research in molecular solvents is well
supported by existing evidence from computer
simulations \cite{Hwang:87,Kuharski:88,Marchi:93,Yelle:97,Hartnig:01}.
The direct consequence of the LRA are the following relations for the
moments of the solute-solvent interaction potential $v_{0s}$:
\begin{equation}
  \label{eq:2-2-3}
  -k_{\text{B}}T \langle v_{0s} \rangle = \langle\left(\delta v_{0s}\right)^2 \rangle = \langle\left(\delta v_{0s}\right)^2 \rangle_0 .
\end{equation}
When the solute-solvent interaction is limited to the coupling of the
solute charges to the solvent dipolar polarization,
$v_{0s}=\langle\Psi|H_{0s}|\Psi\rangle$ in Eq.\ (\ref{eq:1-4}).

The independence of the response function with respect to the solute charge is
propagated into equality of the variance of $v_{0s}$ in equilibrium
with fully charged solute, $\langle\dots\rangle$, and in equilibrium with
uncharged solute, $\langle\dots\rangle_0$.  Figure \ref{fig:3poly} shows the results
of simulations from Ref.\ \onlinecite{DMjcp2:04} for a model diatomic
donor-acceptor complex D--A in a dense solvent of hard sphere point
dipoles. The system is designed to mimic the charge separation, D--A
$\to$ D$^{+}$--A$^{-}$, and charge recombination, D$^{+}$--A$^{-}$ $\to$
D--A, reactions.  The reorganization energies for charge separation,
$\lambda_1=\langle\left(\delta v_{0s}\right)^2\rangle_0/2k_{\text{B}}T$, and for charge recombination, $\lambda_2=\langle \left(\delta
  v_{0s}\right)^2\rangle/2k_{\text{B}}T$, turn out to be very similar over a
broad range of solvent polarities monitored by the dipolar density
parameter $y=(4\pi/9)m^2\rho/k_{\text{B}}T$; $\rho$ is the solvent number
density, $m$ is the solvent molecule permanent dipole moment.

The inhomogeneous character of the response functions is retained
after transformation to $\mathbf{k}$-space. The
function $\bm{\chi}(\mathbf{k},\mathbf{k}')$ then depends on two
wavevectors in contrast to the dependence on a single wave-vector
for the homogeneous dielectric response.  The calculation of
$\bm{\chi}(\mathbf{k},\mathbf{k}')$ is still a major challenge for
microscopic theories of polar solvation. Despite some very active
research in this area for the last 80 years since the formulation of
the Born model for solvation of spherical ions \cite{Born:20}, no
microscopic solution applicable to solutes of arbitrary shape has been
presented so far.  A promising strategy, adopted already in the Born
\cite{Born:20} and Onsager \cite{Onsager:36} models, is to calculate
the response functions in terms of properties of the pure solvent.
This connection can be achieved by considering the polarization
correlation function in the presence of the repulsive core of the
solute [Eq.\ (\ref{eq:2-2-2})]. 

The exclusion of the polarization field from the solute volume is
provided by the Li-Kardar-Chandler approach \cite{Li:92,Chandler:93},
in which the trajectories defining the response function in its path
integral representation are restricted from entering the solute.  The
result of this procedure is an integral equation relating
$\bm{\chi}(\mathbf{k},\mathbf{k}')$ to the non-local susceptibility of
the pure solvent $\bm{\chi}_s(\mathbf{k})$ (with a single
$\mathbf{k}$-vector for the homogeneous response) and the shape of the
solute. The equation for the response function is then equivalent to
the Ornstein-Zernike equation for the solute-solvent correlation
function with the Percus-Yevick closure for the solute-solvent direct
correlation function \cite{Chandler:93}.

No general solution for $\bm{\chi}(\mathbf{k},\mathbf{k}')$ in 
the Li-Kardar-Chandler integral equation has
been obtained so far. One can, however, employ analytical properties
of the response functions to obtain the solvation
chemical potential \cite{DMjcp1:04}
\begin{equation}
  \label{eq:2-2-4}
  - \mu_{0s} = \frac{1}{2}\int\frac{d\mathbf{k}d\mathbf{k}'}{(2\pi)^6} 
  \mathbf{\tilde E}_0(\mathbf{k})\cdot\bm{\chi}(\mathbf{k},\mathbf{k}')\cdot
  \mathbf{\tilde E}_0(-\mathbf{k}') .
\end{equation}
The closed-form result for $\mu_{0s}$ exists when the Fourier transform
of the electric field $\mathbf{\tilde E}_0(\mathbf{k})$ is known in
analytical functional form. This is not the case for many real
problems, when the distribution of molecular charge is given from
force fields or quantum calculations and the Fourier transform of the
field is calculated numerically.  Unfortunately, the analytical
solution is given by the difference of two large numbers almost
canceling each other.  It therefore becomes not very practical in
strongly polar solvents because of accumulation of numerical errors.
To facilitate numerical applications, a mean-field solution for
$\bm{\chi}(\mathbf{k},\mathbf{k}')$ was offered in Ref.\ 
\onlinecite{DMjcp2:04}.  This solution eliminates the inhomogeneous
character of the response function by a non-local renormalization of
its transverse component:
\begin{equation}
  \label{eq:2-4-1}
  \bm{\chi}(\mathbf{k},\mathbf{k}') = (2\pi)^3\delta(\mathbf{k}-\mathbf{k}') \left[\chi^L(\mathbf{k})\mathbf{J}^L  
                                   + \chi^T(\mathbf{k})\mathbf{J}^T\right] ,
\end{equation}
where $\mathbf{J}^L= \mathbf{\hat k}\mathbf{\hat k}$ and $\mathbf{J}^T
= \mathbf{1} - \mathbf{\hat k}\mathbf{\hat k}$ are, respectively, the
longitudinal and transverse projections of a 2-rank tensor with the
axial symmetry established by the direction of the wavevector,
$\mathbf{\hat k}=\mathbf{k}/k$.  The 6D integral of Eq.\ 
(\ref{eq:2-2-4}) is then reduced to the computationally
tractable 3D integral.  

The transverse, $ \chi^T(\mathbf{k})$, and longitudinal,
$\chi^L(\mathbf{k})$, projections in Eq.\ (\ref{eq:2-4-1}) are related to
corresponding components of the susceptibility of the pure polar
solvent
\begin{equation}
  \label{eq:2-4-2}
   \chi^T(\mathbf{k}) =  \chi_s^T(k) \frac{\chi_s^L(0)}{\chi_{\text{tr}}} - 
   f_s\chi_s^L(k) \frac{\mathbf{F}_0\cdot\mathbf{J}^L\cdot\mathbf{\tilde E}_0(\mathbf{k})}
   {\mathbf{F}_0\cdot\mathbf{J}^T\cdot\mathbf{\tilde E}_0(\mathbf{k}) }
\end{equation}
and
\begin{equation}
  \label{eq:2-4-3}
  \chi^L(\mathbf{k}) = \chi_s^L(k) .
\end{equation}
In Eq.\ (\ref{eq:2-4-2}), $\chi_{\text{tr}}=(1/3)\mathrm{Tr}[\bm{\chi}_s(0)]$ and
\begin{equation}
  \label{eq:2-4-4}
    f_s = \frac{2[\chi_s^T(0) - \chi_s^L(0)]}{3\chi_{\text{tr}}}  .
\end{equation}
Further, $\mathbf{\tilde E}_0(\mathbf{k})$ denotes the Fourier
transform of the electric field of the solute calculated on the
volume of the solvent $\Omega$ obtained by excluding the hard repulsive
core of the solute from the solvent
\begin{equation}
  \label{eq:2-4-5}
  \mathbf{\tilde E}_{0}(\mathbf{k}) = \int_{\Omega} \mathbf{E}_{0}(\mathbf{r}) 
   e^{i\mathbf{k}\cdot\mathbf{r}}d\mathbf{r} .
\end{equation}

The mean-field approximation adopted in deriving Eqs.\ 
(\ref{eq:2-4-1})--(\ref{eq:2-4-5}) consists of replacing a generally
non-uniform field of the solvent within the solute by its spatial
average $\mathbf{F}_0$ [Eq.\ (\ref{eq:2-4-2})]. The neglect of the
gradients of the field induced by the solvent within the solute
amounts to taking the dipolar projection of the solute field according
to the following relation:
\begin{equation}
  \label{eq:2-4-6}
  \mathbf{F}_{0} = \ 
        \int_{\Omega}\mathbf{E}_{0}(\mathbf{r})\cdot\mathbf{D}_{\mathbf{r}}\frac{d\mathbf{r}}{r^3} ,
\end{equation}   
where 
\begin{equation}
\label{D}
\mathbf{D}_{\mathbf{r}} = 3\mathbf{\hat r}\mathbf{\hat r} - \mathbf{1}.
\end{equation}
is the dipolar tensor. The electric field $\mathbf{F}_0$ is a
generalization of the Onsager reaction field for the case of
non-spherical solutes with non-dipolar charge distribution.
$\mathbf{F}_0$ reduces to the Onsager field for spherical dipolar
solutes.

The mean-field renormalization of the transverse component of the
response function in Eq.\ (\ref{eq:2-4-2}) resolves the fundamental
difficulty of microscopic solvation theories arising from the fact
that the short-range repulsive perturbation caused by the solute
produces a major change in the polarization response functions
compared to those of the pure solvent.  For instance, a direct
replacement of $\bm{\chi}(\mathbf{k})$ with $\bm{\chi}_{s}(\mathbf{k})$ in
the homogeneous approximation (see Ref.\ \onlinecite{Raineri:94} for
discussion) results in divergent behavior of $\lambda_s$ with increasing
solvent dipole moment \cite{DMcp:96}. The divergence arises from the
transverse component of the response (``transverse catastrophe'')
which has to be included once the dielectric cavity does not coincide
with an equipotential surface of the solute charge distribution
\cite{KKV:76}.  In continuum calculations, the divergent behavior is
eliminated by imposing boundary conditions at the dielectric cavity on
the solution of the Poisson equation. 

Although the problem with the transverse response has long been
recognized in the literature \cite{KKV:76,DMcp:96,Kuznetsov:96}, many
microscopic formulations of solvation thermodynamics and dynamics have
avoided the problem by neglecting the transverse response
\cite{Chandra:89,Bagchi:89,Fried:90} which is also neglected in some
continuum calculations, e.g.\ the Generalized Born approximation
\cite{Schaefer:96}.  Equations (\ref{eq:2-4-1})--(\ref{eq:2-4-5})
provide a general solution of the problem which agrees well with
available simulations of polar solvation \cite{DMjcp1:04,DMjcp2:04}
and experiment on solvation dynamics \cite{DMjcp1:05}. The formalism
is based on the homogeneous solvent susceptibility as input and, once
the susceptibility is defined from computer experiment or liquid-state
theories, can be applied to solvation in an arbitrary isotropic
dielectric.

\begin{figure}[htbp]
  \centering \includegraphics*[width=4cm]{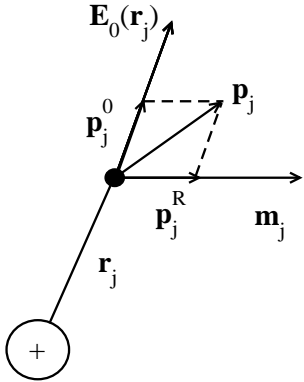}
  \caption{Components of the induced dipole moment at solvent molecule $j$. 
    $\mathbf{p}_j^0$ is produced by the external electric field
    $\mathbf{E}_0(\mathbf{r}_j)$; $\mathbf{p}_j^R$ is produced by the
    reaction field induced in the solvent by the permanent dipole
    $\mathbf{m}_j$.  }
  \label{fig:5poly}
\end{figure}

\subsection{Polarization structure factors}
\label{sec:2-2}
The dipole moment at a given molecule $j$ in a polar-polarizable
solvent is a sum of the permanent dipole $\mathbf{m}_j$ and the
induced dipole $\mathbf{p}_j$ 
\begin{equation}
  \label{eq:10-1}
  \bm{\mu}_j = \mathbf{m}_j + \mathbf{p}_j .
\end{equation}
The total induced dipole then splits into $\mathbf{p}_j^0$
created by the external electric field $\mathbf{E}_0(\mathbf{r}_j)$
and $\mathbf{p}_j^R$ induced by the reaction field
(superscript ``R'') caused by the dipole $\mathbf{m}_j$ itself (Fig.\ 
\ref{fig:5poly}):
\begin{equation}
  \label{eq:10-1-1}
  \mathbf{p}_j = \mathbf{p}_j^0 + \mathbf{p}_j^R 
\end{equation}
The reaction field caused by the dipole $\mathbf{m}_j$ relaxes on the
time-scale of translational-rotational motion of molecule $j$.
Therefore, the induced dipole $\mathbf{p}_j^R$, which follows
adiabatically the reaction field, should be attributed \cite{comPekar}
to the slow nuclear polarization of the solvent $\mathbf{P}_n$. In
contrast, the component $\mathbf{p}_j^0$, following adiabatically the
external field, is attributed to the the fast solvent polarization
$\mathbf{P}_e$. The sum of the permanent dipole $\mathbf{m}_j$ and the
induced dipole $\mathbf{p}_j^R$ makes the effective condensed-phase
dipole \cite{SPH:81}
\begin{equation}
  \label{eq:10-2}
  \mathbf{m}_j' = \mathbf{m}_j + \mathbf{p}_j^R = m' \mathbf{\hat e}_j,
\end{equation}
where $\mathbf{\hat e}_j$ is the unit vector along the direction of
$\mathbf{m}_j$. The dipole moment $m'$ in principle depends on the
instantaneous configuration of the liquid. However, we will not
consider fluctuations of $m'$ here and, following self-consistent
models of polarizable liquids \cite{SPH:81}, will replace $m'$ with its
statistical average value.

The attribution of the electronic polarization in equilibrium with the
electric field of the permanent dipoles to the nuclear (slow)
polarization of the solvent is an essential part of the Pekar
partitioning of the solvent polarization into fast and slow components
\cite{Pekar:46,Pekar:63}. Other partitioning schemes have been
proposed \cite{Brady:85}, but they all lead to the same value of the
solvation energy when correctly implemented \cite{Aguilar:01}.
Computer simulation protocols in which the induced polarization is
self-consistently adjusted to the instantaneous nuclear configuration
provide direct access to the slow polarization in Pekar's definition
\cite{DMjcp:05unpubl}. Self-consistent simulations of polarizable
solvents are used here to test the analytical procedure employed for
the response functions of the nuclear polarization (Sec.\ 
\ref{sec:4-1}).

\begin{figure}[htbp]
  \centering
  \includegraphics*[width=6cm]{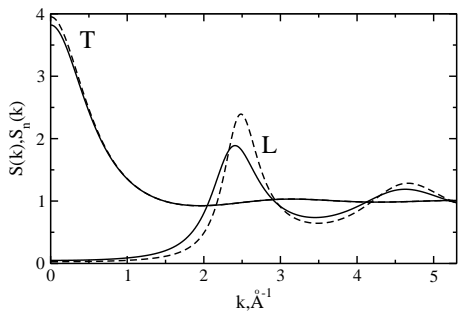}
  \caption{Longitudinal (L) and transverse (T) polarization structure factors
    calculated by using the PPSF with the parameters of ambient water.
    The solid lines refer to the total structure factors $S^{L,T}(k)$,
    the dashed lines refer to the nuclear structure factors
    $S_n^{L,T}(k)$.}
  \label{fig:4poly}
\end{figure}

The total dipolar response function of the homogeneous solvent is a
2-rank tensor describing correlations of dipole moments $\bm{\mu}_j$:
\begin{equation}
  \label{eq:10-4}
  \bm{\chi}_s(\mathbf{k}) = (\beta/ \Omega) \left \langle\sum_{j,k} \bm{\mu}_j\bm{\mu}_k e^{i\mathbf{k}\cdot\mathbf{r}_{jk}} \right \rangle ,
\end{equation}
where $\mathbf{r}_{jk}=\mathbf{r}_j -\mathbf{r}_k$ and brackets refer
to an ensemble average.  Because of the isotropic symmetry of the
solvent, $\bm{\chi}_s(\mathbf{k})$ splits into longitudinal and
transverse components \cite{Madden:84}
\begin{equation}
  \label{eq:10-5}
  \bm{\chi}_s(\mathbf{k}) = \chi_s^L(k) \mathbf{J}^L + \chi_s^T(k)\mathbf{J}^T .
\end{equation}
It is convenient to factor the response function into the effective
density of dipoles $y_{\text{eff}}$, which is mostly affected by the
magnitude of the solvent dipole, and the structure factor, which
reflects dipolar correlations and can be expressed through angular
projections of the pair distribution function \cite{DMjcp2:04}
\begin{equation}
  \label{eq:10-6}
      \bm{\chi}_s(\mathbf{k}) = \frac{3y_{\text{eff}}}{4\pi}\left[S^L(k) \mathbf{J}^L + 
                    S^T(k) \mathbf{J}^T \right] .
\end{equation}
The structure factors $S^{L,T}(k)$ (Fig.\ \ref{fig:4poly}) are defined
based on the unit vectors $\mathbf{\hat u}_j =\bm{\mu}_j / \mu_j$ in the
direction of the respective total dipole moments
\begin{equation}
  \label{eq:2-6}
  \begin{split}
  S^L(k) = & \frac{3}{N}\left\langle \sum_{i,j}(\mathbf{\hat u}_i\cdot\mathbf{\hat k})
            (\mathbf{\hat k}\cdot\mathbf{\hat u}_j) e^{i\mathbf{k}\cdot\mathbf{r}_{ij}}\right\rangle, \\ 
  S^T(k) =  & \frac{3}{2N}\left\langle \sum_{i,j}\left[(\mathbf{\hat u}_i\cdot\mathbf{\hat u}_j)-
     (\mathbf{\hat u}_i\cdot\mathbf{\hat k})(\mathbf{\hat k}\cdot\mathbf{\hat u}_j)\right]
                          e^{i\mathbf{k}\cdot\mathbf{r}_{ij}}\right\rangle .  
  \end{split}
\end{equation}
The effective dipole density in Eq.\ (\ref{eq:10-6}) is 
\begin{equation}
  \label{eq:2-7}
  y_{\text{eff}} = y_p + (4\pi/3)\rho\alpha,\quad y_p = (4\pi/9) \rho(m')^2/k_{\text{B}}T , 
\end{equation}
where $\alpha$ is the dipolar polarizability.  Only the permanent
dipole moment is renormalized by the mean field of the solvent in the
above equation, which corresponds to Wertheim's 1-RPT
theory \cite{Wertheim:79} (2-RPT theory renormalizes the polarizability
$\alpha$ to $\alpha'$, but the 1-RPT version of the theory is in better
agreement with simulations \cite{DMjpca:04}).

The nuclear response function reflects correlated orientations and
positions of dipoles $\mathbf{m}_j'$:
\begin{equation}
  \label{eq:10-7}
  \bm{\chi}_n(\mathbf{k}) = (\beta/ \Omega) 
      \left \langle\sum_{j,k} \mathbf{m}_j'\mathbf{m}_k' e^{i\mathbf{k}\cdot\mathbf{r}_{jk}} \right \rangle .
\end{equation}
Similarly to Eq.\ (\ref{eq:10-6}), $\bm{\chi}_n(\mathbf{k})$ can be
separated into the longitudinal and transverse components
\begin{equation}
  \label{eq:10-8}
      \bm{\chi}_n(\mathbf{k}) = \frac{3y_{p}}{4\pi}\left[S_n^L(k) \mathbf{J}^L + 
                    S_n^T(k) \mathbf{J}^T \right] .
\end{equation}
The nuclear structure factors are defined by Eq.\ (\ref{eq:2-6}), in
which the unit vectors $\mathbf{\hat u}_j$ are replaced by the unit
vectors $\mathbf{\hat e}_j$ [Eq.\ (\ref{eq:10-2})].

The $k=0$ values of the structure factors are related to the
macroscopic dielectric properties of the solvent. 
The total polarization response is defined through the
static dielectric constant $\epsilon_s$
\begin{equation}
  \label{eq:3-10}
    \begin{split}
    S^L(0) & = \frac{\epsilon_s - 1}{3\epsilon_sy_{\text{eff}}},\\
    S^T(0) & = \frac{\epsilon_s - 1}{3y_{\text{eff}}} .
    \end{split}
\end{equation}
The nuclear structure factors depend, in addition, on the high-frequency dielectric
constant $\epsilon_{\infty}$ \cite{DMjcp:05unpubl}
\begin{equation}
\label{eq:3-11}
   \begin{split} 
    S_n^L(0) & = \frac{c_0}{3y_{p}}, \\
    S_n^T(0) & = \frac{\epsilon_s - \epsilon_{\infty}}{3y_{p}} ,
   \end{split}
\end{equation}
where 
\begin{equation}
  \label{eqPekar}
    c_0=1/ \epsilon_{\infty} - 1/ \epsilon_s   
\end{equation}
is the Pekar factor. 

Both $S_n^{L,T}(k)$ and $S^{L,T}(k)$ tend to unity at $k\to \infty$.  This
limit is the result of the point multipole approximation for the
intramolecular charge distribution within the solvent molecules.  In
contrast, charge-charge structure factors defined on interaction-site
models of liquids decay to zero at $k\to\infty$ (Refs.\ 
\onlinecite{Perng:96,Raineri:99,Perng:99}). The region of $k$-values
where this distinction becomes important is, however, insignificant
for the calculation of solvation thermodynamics (see below).  The
nuclear and the total structure factors differ in the range of small
$k$-values and around the longitudinal peak as a result of the
influence of the high-frequency dielectric constant of the solvent
(Fig.\ \ref{fig:4poly}). The effect of $\epsilon_{\infty}$ on the longitudinal
peak is insignificant for the calculation of the reorganization
energy. Therefore, it is the range of small $k$-values and, in
addition, the dependence of the liquid-state dipole moment $m'$ on the
solvent polarizability, that ultimately determine the variation of the
solvent reorganization energy with the solvent high-frequency
dielectric constant $\epsilon_{\infty}$ (see below).

\subsection{ET thermodynamics}
\label{sec:2-3}
The solvation thermodynamics of ET is determined by the solvent
reorganization energy and the solvent component of the free energy
gap. They are defined in terms of the nuclear and total response
functions by the following relations
\begin{equation}
  \label{eq:2-3}
  \lambda_s = \frac{1}{2} \int \frac{d\mathbf{k}d\mathbf{k}'}{(2\pi)^6} 
       \Delta \mathbf{\tilde E}_0(\mathbf{k})
         \cdot \bm{\chi}_n(\mathbf{k},\mathbf{k}') \cdot  
       \Delta \mathbf{\tilde E}_0(-\mathbf{k}')
\end{equation}
and
\begin{equation}
  \label{eq:2-4}
  \Delta G_s =  - \int \frac{d\mathbf{k}d\mathbf{k}'}{(2\pi)^6} \Delta \mathbf{\tilde E}_0(\mathbf{k})\cdot
            \bm{\chi}(\mathbf{k},\mathbf{k}') \cdot \mathbf{\bar E}_0(-\mathbf{k}') .
\end{equation}
In Eqs.\ (\ref{eq:2-3}) and (\ref{eq:2-4}), $\Delta \mathbf{\tilde
  E}_0(\mathbf{k}) = \mathbf{\tilde E}_{02}(\mathbf{k}) -
\mathbf{\tilde E}_{01}(\mathbf{k})$ and $ \mathbf{\bar
  E}_0(\mathbf{k}) = (\mathbf{\tilde
  E}_{02}(\mathbf{k})+\mathbf{\tilde E}_{01}(\mathbf{k}))/2$;
$\mathbf{\tilde E}_{0i}(\mathbf{k})$ are the Fourier transforms of the
solute electric field in the initial ($i=1$) and final ($i=2$) ET
states taken over the volume $\Omega$ occupied by the solvent [Eq.\ 
(\ref{eq:2-4-5})].

The mean-field solution for the response functions [Eq.\ (\ref{eq:2-4-1})]
splits both the solvent reorganization energy and the free energy gap
into their corresponding longitudinal and transverse components:
\begin{equation}
  \label{eq:2-8}
  \lambda_s = \lambda_s^L + \lambda_s^T
\end{equation}
and
\begin{equation}
  \label{eq:2-9}
  \Delta G_s = \Delta G_s^L + \Delta G_s^T .
\end{equation} 
Each projection is obtained as a $\mathbf{k}$-integral with the
corresponding polarization structure factor. For the ``T'' projections
one gets
\begin{equation}
  \label{eq:2-10}
  \lambda_s^T = \frac{3y_p}{8\pi}\,\frac{S_n^L(0)}{g_{Kn}}\int\frac{d\mathbf{k}}{(2\pi)^3}
          \left|\Delta \tilde E_0^T(\mathbf{k}) \right|^2 S_n^T(k) 
\end{equation}
and
\begin{equation}
  \label{eq:2-11}
  \begin{split}
  \Delta G_s^T & = -\frac{3y_{\text{eff}}}{8\pi}\,\frac{S^L(0)}{g_K}\int\frac{d\mathbf{k}}{(2\pi)^3}\\
       &   \left[|\tilde E_{02}^T(\mathbf{k})|^2 - 
              |\tilde E_{01}^T(\mathbf{k})|^2 \right] S^T(k) .
 \end{split}
\end{equation}
In Eqs.\ (\ref{eq:2-10}) and (\ref{eq:2-11}), 
\begin{equation}
  \label{eqGKn}
  g_{Kn} = (1/3)\left[S_n^L(0) + 2S_n^T(0) \right]
\end{equation}
and
\begin{equation}
  \label{eqGK}
  g_{Kn} = (1/3)\left[S^L(0) + 2S^T(0) \right]
\end{equation}
are the nuclear and total Kirkwood factors, respectively.

The longitudinal components of free energies, $\lambda_s^L$ and $\Delta G_s^L$,
include both the longitudinal and transverse projections of the solute
field:
\begin{equation}
  \label{eq:2-12}
  \lambda_s^L = \frac{3y_p}{8\pi}\int \frac{d\mathbf{k}}{(2\pi)^3}
          \mathcal{E}^{\text{eff}}_{\Delta}(\mathbf{k}) S_n^L(k)
\end{equation}
and
\begin{equation}
  \label{eq:2-13}
  \Delta G_s^L = - \frac{3y_{\text{eff}}}{8\pi}\int \frac{d\mathbf{k}}{(2\pi)^3}
          \left(\mathcal{E}_2^{\text{eff}}(\mathbf{k}) - \mathcal{E}_1^{\text{eff}}(\mathbf{k})\right) S^L(k).
\end{equation}
In Eqs.\ (\ref{eq:2-12}) and (\ref{eq:2-13}), 
\begin{equation}
  \label{eq:2-14}
  \mathcal{E}^{\text{eff}}_{\Delta}(\mathbf{k})= |\Delta \tilde E_0^L(\mathbf{k})|^2 
  - f_{n} |\Delta \tilde E_0^T(\mathbf{k})|^2 
    \frac{\Delta \mathbf{F}_{0}\cdot\mathbf{J}^L\cdot \Delta\mathbf{\tilde E_0}(\mathbf{k}) }
   {\Delta \mathbf{F}_{0}\cdot\mathbf{J}^T\cdot\Delta \mathbf{\tilde E}_0(\mathbf{k})}
\end{equation}
and
\begin{equation}
  \label{eq:2-14-1}
  \mathcal{E}^{\text{eff}}_{i}(\mathbf{k})= |\tilde E_{0i}^L(\mathbf{k})|^2 - 
    f_{s}|\tilde E_{0i}^T(\mathbf{k})|^2 
    \frac{\mathbf{F}_{0i}\cdot\mathbf{J}^L\cdot\mathbf{\tilde E}_{0i}(\mathbf{k})}
   {\mathbf{F}_{0i}\cdot\mathbf{J}^T\cdot \mathbf{\tilde E}_{0i}(\mathbf{k})} .
\end{equation}
The longitudinal and transverse components of the electrostatic energy 
density in Eqs.\ (\ref{eq:2-10})--(\ref{eq:2-13}) are defined as
\begin{equation}
  \label{eq:2-14-2}
   \begin{split}
  |\Delta E_0^{L,T}(\mathbf{k})|^2 &= \Delta \mathbf{\tilde E}_0(\mathbf{k})\cdot \mathbf{J}^{L,T} \cdot
                            \Delta \mathbf{\tilde E}_0(-\mathbf{k}),\\
    |E_{0i}^{L,T}(\mathbf{k})|^2 & = \mathbf{\tilde E}_{0i}(\mathbf{k}) \cdot \mathbf{J}^{L,T} \cdot
                            \mathbf{\tilde E}_{0i}(-\mathbf{k}) .
   \end{split}
\end{equation}
The effective fields $\mathcal{E}^{\text{eff}}_{\Delta}(\mathbf{k})$ and
$\mathcal{E}^{\text{eff}}_{i}(\mathbf{k})$ depend on the symmetry of
the charge distribution within the solute analogously to the result of
imposing the boundary conditions on the solution of the Poisson
equation in continuum electrostatics.

The electric field $\mathbf{F}_{0i}$ in Eq.\ (\ref{eq:2-14-1}) is a
generalization of the Onsager reaction cavity field \cite{Onsager:36}
to the case of solutes of non-spherical shape and non-point-dipole
charge distribution.  This field is obtained by summing up a
continuous distribution of dipolar electric fields induced by the
solute in the solvent volume:
\begin{equation}
  \label{eq:2-15}
  \mathbf{F}_{0i} = \ 
        \int_{\Omega}\mathbf{E}_{0i}(\mathbf{r})\cdot\mathbf{D}_{\mathbf{r}}\frac{d\mathbf{r}}{r^3} ,
\end{equation}
where $\mathbf{D}_{\mathbf{r}}$ is given by Eq.\ (\ref{D}).  Also, $\Delta
\mathbf{F}_0$ in Eq.\ (\ref{eq:2-14}) is $\Delta \mathbf{F}_0 =
\mathbf{F}_{02} - \mathbf{F}_{01}$. $\mathbf{F}_{0i}$ becomes the
standard Onsager reaction field for a point dipole at the center of a
spherical cavity. Finally, in Eqs.\ (\ref{eq:2-14}) and
(\ref{eq:2-14-1}),
\begin{equation}
  \label{eq:2-16-1}
   f_s=\frac{2(\epsilon_s -1)}{2\epsilon_s + 1} 
\end{equation}
is the usual Onsager polarity parameter \cite{Onsager:36} and the
corresponding polarity parameter for the nuclear polarization is
\begin{equation}
  \label{eq:2-16}
   f_n=\frac{2(\epsilon_{\infty}\epsilon_s-1)}{2\epsilon_{\infty}\epsilon_s + 1} . 
\end{equation}

\section{Calculation procedure}
\label{sec:4}
\begin{figure}[htbp]
  \centering
  \includegraphics*[width=6cm]{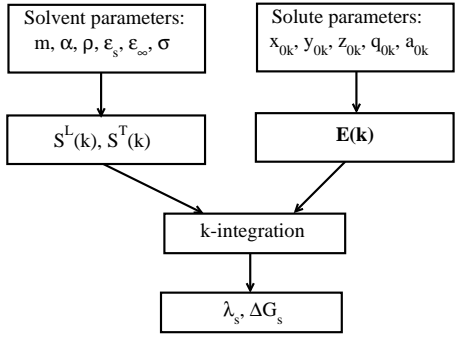}
  \caption{Diagram of the calculation algorithm. Solvent parameters include:
    $m$ (gas-phase dipole moment), $\alpha$ (gas-phase dipolar
    polarizability), $\epsilon_{\infty}$ (high-frequency dielectric constant),
    $\epsilon_s$ (static dielectric constant), and $\sigma$ (effective hard sphere
    diameter of the solvent molecules). Parameters
    $x_{0k}$, $y_{0k}$, $z_{0k}$ stand for Cartesian coordinated of the solute
    atoms, $q_{0k}$ are partial charges, and $a_{0k}$ are atomic vdW radii.
  }
  \label{fig:7poly}
\end{figure}

The formalism outlined above is realized in a computational algorithm
sketched in Figure \ref{fig:7poly}. It includes two branches, one is for
the solvent part of the calculation and another is for the solute
part.  The two parts are combined together in the integration over the
inverted space, which yields the reorganization energy ($\lambda_s$) and the
total free energy of nuclear plus electronic solvation ($\Delta G_s$). We
start with describing the solvent branch followed by the outline of
the solute part.

\subsection{Solvent}
\label{sec:4-1}
The calculation of the structure factors in the solvent branch in
Fig.\ \ref{fig:7poly} requires a set of experimental input parameters: $m$
(gas-phase dipole moment), $\alpha$ (gas-phase dipolar polarizability),
$\epsilon_{\infty}$ (high-frequency dielectric constant), $\epsilon_s$ (static dielectric
constant), and $\sigma$ (effective hard sphere diameter of the solvent
molecules).  The hard sphere diameter is obtained from the experimental
compressibility of the solvent by fitting it to the compressibility
found from the generalized van der Waals (vdW) equation of state
\cite{DMjpc:95}. Based on these parameters, an analytical procedure
has been recently proposed to calculate $S^{L,T}(k)$ \cite{DMjcp2:04}. This
parameterization, called parametrized polarization structure factors
(PPSF), makes use of the analytical solution of the mean-spherical
approximation (MSA) for dipolar fluids \cite{Wertheim:71}. The MSA
solution gives $S^{L,T}(k)$ in terms of the Baxter function $Q(k\sigma,\eta)$
appearing as solution of Percus-Yevick integral equations for hard
sphere fluids \cite{Gubbins:84}
\begin{equation}
  \label{eq:3-6}
  S(k\sigma,\eta) = |Q(k\sigma,\eta)|^{-2} ,
\end{equation}
where
  \begin{equation}
  \label{eq:3-7}
  \begin{split}
  Q(k\sigma,\eta) = &1 - 12 \eta \int_0^1e^{ik\sigma t} \\
           & \left[a(\eta)(t^2-1)/2 - b(\eta)(t-1)\right]dt
  \end{split}
\end{equation}
and $a(\eta)=(1+2\eta)/(1-\eta)^2$, $b(\eta)=-3\eta/2(1-\eta)^2$.  For a fluid of hard
sphere molecules, $\eta=(\pi/6)\rho\sigma^3$ is the packing density, equal to the ratio of the
volume of the solvent molecules to the volume of the liquid. In the
MSA, the $S^{L,T}(k)$ are obtained by setting $\eta=2\xi$ for $S^{L}(k)$
and $\eta=-\xi$ for $S^{T}(k)$ in Eq.\ (\ref{eq:3-6}). Here, $\xi$ is the MSA
polarity parameter which can be related either to the dipolar density
$y_{\text{eff}}$ or to the static dielectric constant $\epsilon_s$
\cite{Wertheim:71}.

Two problems arise when dealing with the reorganization energy
calculations using the polarization structure factors from the MSA.
First, one needs a general procedure which would provide the nuclear
structure factors $S_n^{L,T}(k)$ in polarizable solvents in contrast
to total structure factors $S^{L,T}(k)$ given by the MSA solution.
Such a formalism should thus exclude (quantum) fluctuations of the
induced solvent dipoles $\mathbf{p}_j^0$ which are not included in the
nuclear polarization field (Fig.\ \ref{fig:5poly}). Second, the MSA
does not give a consistent description of the dielectric properties of
polar solvents, i.e.\ the polarity parameters $\xi$ calculated from
$y_{\text{eff}}$ and $\epsilon_s$ are quite different. The PPSF procedure
goes around the second problem by considering $y_{\text{eff}}$ and
$\epsilon_s$ as two independent input parameters used to calculate
$S^{L,T}(k)$.  A convenient way to introduce the two-parameter scheme
is to specify two separate polarity parameters which are obtained from
the longitudinal and transverse structure factors at $k=0$:
\begin{equation}
  \label{eq:3-8}
  \begin{split}
     \frac{(1-2\xi^L)^4}{(1+4\xi^L)^2}  = & S^L(0),\\
     \frac{(1+\xi^T)^4}{(1-2\xi^T)^2}   = & S^T(0).
  \end{split}
\end{equation}

\begin{figure}[htbp]
  \centering \includegraphics*[width=6cm]{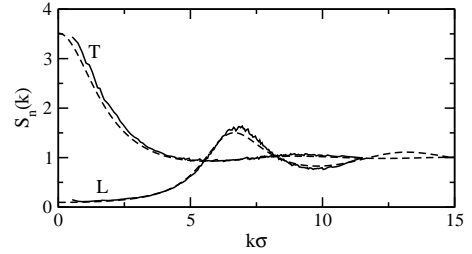}
  \caption{Nuclear longitudinal (L) and transverse (T) structure factors 
    from the PPSF (dashed lines) and MC simulations (solid lines).  MC
    simulations are carried out for a fluid of 1372 hard spheres with
    permanent dipole $m$, diameter $\sigma$, polarizability $\alpha$, and
    density $\rho$: $(m^*)^2=\beta m^2/ \sigma^3=1.0$, $\alpha^*=\alpha/ \sigma^3=0.06$, $\rho\sigma^3 =
    0.8$. The dielectric properties from the simulations are:
    $\epsilon_s=21.4$,  $y_{\text{eff}}=1.57$, and $y_p=1.54$; $\epsilon_{\infty}=1.75$ is obtained from
    the Clausius-Mossotti equation. 
  }
  \label{fig:8poly}
\end{figure}

Separate definitions of $\xi^{L}$ and $\xi^T$ in terms of $S^L(0)$ and
$S^T(0)$ [Eq.\ (\ref{eq:3-10})] allows us to incorporate contributions
to macroscopic dielectric properties which are not present in the
model of dipolar HS fluids. Specifically, the magnitude of parameter
$y_{\text{eff}}$, calculated according to Wertheim's 1-RPT algorithm
\cite{Wertheim:79}, defines the solvent dipolar strength  which
strongly affects the dielectric constant. However, $\epsilon_s$ also depends
on such factors as solvent quadrupolar moment \cite{SPH:81}, solvent
non-sphericity, etc.  The influence of these factors is incorporated
into $S^{L,T}(0)$ through the dielectric constant.  Similarly, the
polarity parameters $\xi_n^L$ and $\xi_n^T$ are calculated from Eq.\ 
(\ref{eq:3-8}) with $S^{L,T}(0)$ replaced by $S_n^{L,T}(0)$ taken from
Eq.\ (\ref{eq:3-11}).

Dipolar projections of the structure factors of molecular liquids
modeled by site-site interaction potentials have been studied
previously \cite{Fonseca:90,Raineri:93,Skaf:95,Perng:99}. The PPSF
procedure has also been tested against MC simulations of dipolar hard
sphere fluids \cite{DMjcp2:04}. However, the structure factors arising
from the nuclear polarization as well as the applicability of the PPSF to
non-spherical molecules with site-site potentials have not been
previously tested. This is the aim of the Monte Carlo (MC) and MD
simulations carried out in this study.  The details of the simulation
protocol are given in Appendix \ref{A1} and here we focus only on the
results.

Figure \ref{fig:8poly} shows the comparison of the transverse and
longitudinal components of the nuclear structure factors calculated
from the PPSF and from MC simulations.  The MC simulations
(dashed lines in Fig.\ \ref{fig:8poly}) have been performed on a fluid
of 1372 polarizable dipolar hard spheres characterized by dipole
moment $m$, diameter $\sigma$, and isotropic polarizability $\alpha$ ($(m^*)^2=\beta
m^2/ \sigma^3=1.0$, $\alpha^*=\alpha/ \sigma^3=0.06$, Appendix \ref{A1}). Since the
simulation protocol generates the induced polarization in equilibrium
with the nuclear configuration of the solvent \cite{DMjpca:04}, the
generated ensamble yields the nuclear polarization in the Pekar partitioning
\cite{Pekar:63}. 

The PPSF nuclear structure factors are calculated by the relations:
\begin{equation}
  \label{eq:3-13}
 S_n^T(k) =   |Q(k\sigma,-\xi_n^T)|^{-2} 
\end{equation}
and 
\begin{equation}
  \label{eq:3-14}
 S_n^L(k) =   |Q(\kappa k\sigma,2\xi_n^L)|^{-2}  .
\end{equation}
In Eq.\ (\ref{eq:3-14}), $\kappa=0.95$ is an empirical parameter introduced
for a better agreement between the PPSF and MC simulations of
non-polarizable dipolar fluids \cite{DMjcp:05unpubl}. The simulations
and the PPSF agree well in the entire range of solvent
polarizabilities $\alpha^*=\alpha/ \sigma^3=0.01-0.08$ studied by simulations
\cite{DMjcp:05unpubl}.

\begin{figure}[htbp]
  \centering
 \includegraphics*[width=6cm]{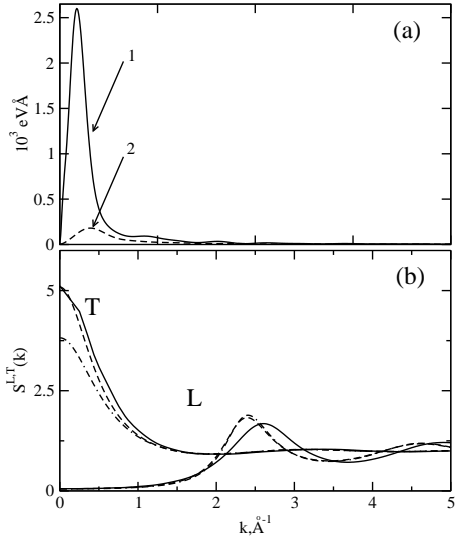}  
  \caption{Upper panel: longitudinal (1), 
    $k^2\langle\mathcal{E}_{\Delta}^{\text{eff}}(\mathbf{k}))\rangle_{\mathbf{\hat
        k}}$, and transverse (2), $k^2\langle(\Delta E^T_0(k))^2\rangle_{\mathbf{\hat
        k}}$ components of the electrostatic energy density of complex
    \textbf{1} entering the $k$-integral in Eqs.\ (\ref{eq:2-12}) and
    (\ref{eq:2-10}), respectively. $\langle\dots \rangle_{\mathbf{\hat k}}$
    denotes the average over the orientations of the wavevector
    $\mathbf{k}$.  Lower panel: longitudinal (L) and transverse (T)
    structure factors for TIP3P water at 298 K.  The solid lines refer
    to the results of MD simulations.  Dashed lines indicate the
    results of PPSF calculations with the input parameters
    corresponding to the TIP3P force field (Table \ref{tab:22},
    $m=2.35$ D, $\epsilon_s=95.4$, $\epsilon_{\infty}=1.0$) and $\sigma=2.87$ \AA.  The dash-dotted lines refer to the 
    nuclear structure factors of ambient water. The graphs
    in the upper and lower panels are plotted against the same scale
    of $k$-values to indicate that the details of the structure
    factors beyond approximately $k\sigma\simeq\pi$ are insignificant for the
    calculation of the reorganization energy.}
  \label{fig:9poly}
\end{figure}

The MSA solution in Eq.\ (\ref{eq:3-6}) was derived for a model liquid
of dipolar hard spheres. The parameterization introduced by the PPSF suggests
to use the experimental $\epsilon_s$ to accommodate empirically the features
which are not included in the MSA solution. Two factors, often present
in real polar solvents, molecular quadrupoles and non-sphericity, are
expected to affect significantly the form of the structure factors.
Therefore, we have performed MD simulations for two solvents with
well-developed force fields, water \cite{jorg:83} and acetonitrile
\cite{acnMadden:84}. Water is a relatively symmetric molecule with a
very large quadrupole moment $Q$ \cite{comQ}  ($(Q^*)^2 =\beta Q^2 / \sigma^5 = 1.1$) among
commonly used molecular solvents. On the other hand, acetonitrile has
a small quadrupole moment ($(Q^*)^2=0.13$), but the molecule is very
non-spherical with the aspect ratio $\simeq 3$.  Therefore, these two
extreme cases may provide a good test of the ability of the PPSF to
incorporate the complications related to molecular specifics of the
solvents in terms of their macroscopic dielectric constants.

\begin{figure}[htbp]
  \centering
  \includegraphics*[width=6cm]{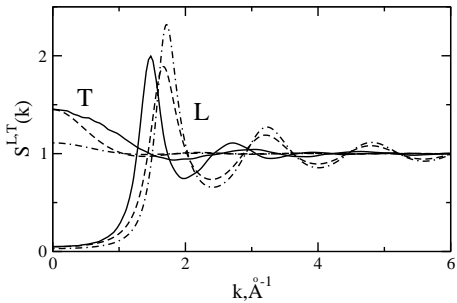} 
  \caption{Longitudinal (L) and transverse (T) polarization 
    structure factors of acetonitrile at 298 K. The solid lines refer
    to the results of MD simulations.  Dashed lines indicate the
    results of PPSF calculations with the input parameters
    corresponding to ACN3 (Table \ref{tab:22}, $m=4.12$ D,
    $\epsilon_s=29.6$, $\epsilon_{\infty}=1.0$) and $\sigma=4.14$ \AA.
    The dash-doted lines indicate the nuclear structure factors
    $S_n^{L,T}(k)$ from the PPSF with the parameters of ambient
    acetonitrile: $m=3.9$ D, $\epsilon_s=35.9$,
    $\epsilon_{\infty}=1.8$, $\alpha=4.48$ \AA$^3$, $\eta=0.424$,
    $\sigma=4.14$ \AA. }
  \label{fig:10poly}
\end{figure}

Figure \ref{fig:9poly} (lower panel) shows the comparison of the
simulation results for TIP3P water to the PPSF.  A slightly wrong
positioning of the longitudinal peak may be related to a different
hard sphere diameter of TIP3P water (see Table \ref{tab:5} in Appendix
\ref{A1}) compared to the hard sphere diameter of water at ambient
conditions used in scaling wavevectors in Figure \ref{fig:9poly}.  A
downward scaling of $\sigma$ by just 5\% results in a very good match
between calculated and simulated structure factors. As expected, the
steric effects of packing the solvent molecules in dense liquids is
the main factor determining the position of the longitudinal peak.
This indeed is seen in Fig.\ \ref{fig:10poly} for simulations of
acetonitrile.  The effective hard sphere diameter obtained from
solvent compressibility does not accommodate the fact that linear
dipoles tend to pack side-to-side pointing in opposite directions.
The longitudinal thus peak effectively reflects a lower
molecular diameter. The preferential opposite orientation of the
dipoles leads to a low Kirkwood factor and the dielectric constant
much lower than one would expect for a dipolar solvent with such large
dipole moment ($4.12$ D for the force field by Edwards, Madden, and
McDonald \cite{acnMadden:84}). As a result, the transverse structure
factor does not change with $k$ as much as it does for hard sphere
dipolar liquids (cf.\ Figs.\ \ref{fig:8poly} and \ref{fig:9poly} to Fig.\ 
\ref{fig:10poly}).  As is seen, the PPSF with $\epsilon_s$ from MD simulations
accommodates this feature of the solvent quite well.

Figure \ref{fig:9poly} compares on the common scale the $k$-dependence of
the longitudinal and transverse components of the electrostatic energy
density of complex \textbf{1},
$k^2\langle\mathcal{E}_{\Delta}^{\text{eff}}(\mathbf{k})\rangle_{\mathbf{\hat k}}$ and
$k^2\langle |\Delta E_0^T(\mathbf{k})|^2\rangle_{\mathbf{\hat k}}$, with the
longitudinal and transverse components of the polarization structure
factors ($\langle\dots\rangle_{\mathbf{\hat k}}$ refers to the average over the
orientations of the wavevector $\mathbf{k}$). This comparison shows
that details of the molecular structure of the polar solvent affecting
the range of $k$-values beyond the limit of $k\simeq \pi/ \sigma$ are
insignificant for the calculation of the reorganization energy and the
free energy gap.  Therefore, the discrepancies in the position of the
longitudinal peak between the simulations and the PPSF do not
noticeably affect the results of calculations.  This statement also
applies to the range of $k$-values ($k>2\pi/l_s$, where $l_s$ is the
characteristic distance between partial charges within the solvent
molecule) at which the multipolar approximation for the charge
distribution within the solvent molecules breaks down. The
charge-charge structure factors calculated on site-site interaction
potentials
\cite{Bopp:96,Perng:96,Skaf:97,Bopp:98,Omelyan:99,Raineri:99,Perng:99}
then decay to zero instead of approaching the unity limit
($S^{L,T}(k)\to 1$ at $k\to\infty$) of multipolar approximations
\cite{Fonseca:90,Skaf:95,Bopp:98}. The range of $k$-values where the
inaccuracy of the multipolar approximation becomes significant lays
beyond the range of small $k$-values affecting the calculation of
thermodynamic properties unless the solute is much smaller than the
solvent.

\subsection{Solute}
\label{sec:4-2}
The solute branch of the calculation algorithm (Fig.\ \ref{fig:7poly})
consists of the numerical calculation of the Fourier transform of the
electric field outside the solute placed in the vacuum. The
direct-space electric fields in the initial and final states of the
solute are given by a superposition of electric fields produced by
partial charges $q_{0k}^i$
\begin{equation}
  \label{eq:3-1}
 \mathbf{E}_{0i}(\mathbf{r}) = \sum_{k=1}^{M_0} q_{0k}^i \frac{\mathbf{r}-\mathbf{r}_{0k}}
                                    {|\mathbf{r}-\mathbf{r}_{0k}|^3} , 
\end{equation} 
where the sum runs over $M_0$ partial charges localized on solute atoms.
The field $\mathbf{E}_{0i}(\mathbf{r})$ is Fourier transformed in the
region $\Omega$ accessible to the solvent molecules [Eq.\ (\ref{eq:2-4-5})].
The region $\Omega$ is generated by assigning vdW radii to all atoms of the
solute and then adding the hard sphere radius $\sigma /2$ of the solvent ($\sigma=2.87$ \AA{}
for water and 4.14 \AA{} for acetonitrile).  This creates the
solvent-accessible surface (SAS).  The definition of the solute field
thus requires atomic coordinates and vdW radii of $N_0$ atoms of the
solute and $M_0$ partial charges $q_{0k}$ to be used in Eq.\ 
(\ref{eq:3-1}) (indicated as $x_{0k}$, $y_{0k}$, $z_{0k}$, $q_{0k}$ in Fig.\ 
\ref{fig:7poly}).

The infinite-space Fourier transform of the Coulomb electric field [Eq.\ 
(\ref{eq:2-4-5})] is numerically divergent \cite{DMjcp2:04}. This numerical problem is
obviated by splitting the region of integration into the inner part
between the SAS and a cutoff sphere and the region outside the cutoff
sphere. The Fourier transform within the sphere is calculated
numerically by the Fast Fourier Transform (FFT) technique
\cite{Fortran:96} on a cube with the center at the geometrical center
of the DSA complex
\begin{equation}
  \label{eq:3-3}
  \mathbf{r}_c = N_0^{-1} \sum_{k=1}^{N_0} \mathbf{r}_{0k} .
\end{equation}
The length of the cube is chosen by multiplying the maximum extension
of the molecule measured from $\mathbf{r}_c$ by a factor of 9. This
choice yields a sufficiently small increment of the $\mathbf{k}$-grid
necessary for the inverted-space integration and, at the same time,
avoids numerical errors arising from artificial periodicity imposed by
a finite-size numerical FFT technique. The FFT calculation was done on
a grid of dimension $256\times256\times256$ and the step of 0.5 \AA.  Calculations
on complex \textbf{1} involved 143 atoms holding partial charges
$q_{0k}^i$.  The charge shifts ($\Delta q_k=q_{0k}^2 - q_{0k}^1$) and
coordinates used in the solvent reorganization and free energy
calculations are the same as those reported in Ref.\ \onlinecite{Ungar:99}.
The individual (i.e., initial and final state) charges used in the
reaction free energy calculations are also taken from Ref.\ 
\onlinecite{Ungar:99}.  The field $\mathbf{\tilde E}_{0i}(\mathbf{k})$
obtained by combining the numerical and analytical parts is used to
calculate the longitudinal and transverse components of the
electrostatic energy density in Eqs.\ (\ref{eq:2-14-1}) and
(\ref{eq:2-14-2}).  These components are then used in the
$k$-integrals with the polarization structure factors (Eqs.\ 
(\ref{eq:2-10})--(\ref{eq:2-13}); also see Fig.\ \ref{fig:7poly}).

\begin{table}[htbp]
  \centering
  \caption{{\label{tab:1}}Reorganization energy (kcal/mol) of \textbf{1} in water. 
           All calculations except those in the last column are done with $\epsilon_{\infty}=1.0$. }  
  \begin{tabular}[c]{cccccccccc}
\hline
\hline    
T, K & $\eta$\footnotemark[1] 
     & $y_p$\footnotemark[2] 
     & $\epsilon_s$\footnotemark[3] 
     & $\lambda_p$\footnotemark[4] 
     & $\lambda_p$\footnotemark[5] 
     & $\lambda_p$\footnotemark[6] 
     & $\lambda_{pq}$\footnotemark[6] 
     & $\lambda_{q}$\footnotemark[6] 
     & $\lambda_p$\footnotemark[7] \\ 
\hline
288 & 0.4110 & 6.44 & 107.7 & 64.93 & 39.60 & 64.35 & 1.55 & 4.37 & 45.88 \\
293 & 0.4104 & 6.32 & 102.1 & 64.52 & 39.58 & 64.26 & 1.45 & 4.33 & 45.47 \\
298 & 0.4098 & 6.21 & 97.5  & 64.11 & 39.56 & 63.93 & 1.38 & 4.25 & 45.07 \\
303 & 0.4092 & 6.09 & 96.0  & 63.67 & 39.55 & 63.52 & 1.42 & 4.13 & 44.68 \\
308 & 0.4085 & 5.99 & 93.7  & 63.25 & 39.54 & 62.98 & 1.22 & 4.14 & 44.30 \\
\hline
\hline
  \end{tabular}
\footnotetext[1]{Packing fraction calculated with $\sigma=2.87$ \AA{} and the isobaric
                 expansion coefficient $\alpha_p=2.96\times10^{-4}$ K$^{-1}$. }
\footnotetext[2]{TIP3P water has the permanent dipole of 2.35 D scaled up from 
                 the vacuum dipole moment of water, 1.83 D, to account for
                 the mean-field effect of the induced dipoles. }
\footnotetext[3]{Calculated from MD simulations as described in Appendix A.}
\footnotetext[4]{Calculations with the PPSF structure factors with 
                 $\epsilon_{\infty}=1.0$ and $\epsilon_s$ from MD simulations. $\lambda_p$ stands for the reorganization energy arising from the interaction between the solute electric field and the solvent dipoles, $\lambda_q$ comes from the interaction between the solute field gradient and solvent quadrupoles, $\lambda_{pq}$ is the mixed term from correlated fluctuations
of dipoles and quadrupoles on different solvent molecules, see Eq.\ (\ref{eq:lambda}).}
\footnotetext[5]{Continuum limit $S_n^{L,T}(k)=S^{L,T}(0)$ at 
                 $\epsilon_{\infty}=1.0$ and $\epsilon_s$ from MD simulations.}
\footnotetext[6]{Calculations with the structure factors from MD simulations.}
\footnotetext[7]{Calculations with the PPSF structure factors with 
                 the solvent parameters of ambient water.}
\end{table}

\section{Results and comparison to experiment}
\label{sec:5}
\subsection{Solvent reorganization energy}\label{sec:5-1}
The solvent reorganization energy of complex \textbf{1} was previously
obtained from MD simulations of this complex in TIP3P water
\cite{Ungar:99}. The permanent dipole moment in this force field is
enhanced from the vacuum dipole of 1.87 D to 2.35 D to account for
water polarizability. This results in a dielectric constant of
$\epsilon_s=97.5$ from our simulations, which agrees well with $\epsilon_s=97.0$
found in the literature \cite{Guillot:02}. Table \ref{tab:1} lists the
results of calculations of the reorganization energy with structure
factors from the PPSF (column 5) and from MD simulations (column 7).
The density of the solvent in the $NVT$ simulations was adjusted at
each temperature in order to reproduce the expansivity
$\alpha_p=2.96\times10^{-4}$ K$^{-1}$ of TIP3P water \cite{xxx:20nsSIM}. The
temperature derivative of the reorganization energy thus gives the
constant-pressure reorganization entropy corresponding to conditions
normally employed in experiment,
\begin{equation}
\label{eq:4-1}
S_{\lambda}= -(\partial \lambda_s / \partial T)_P  .
\end{equation}
Overall, there is an exceptionally good agreement between the
reorganization energies calculated by using the structure factors from
PPSF and MD simulations.  This is not surprising in view of the very
good agreement between the two sets of structure factors shown in
Fig.\ \ref{fig:9poly}. 

The PPSF result at 298 K, $\lambda_s=64.11$ kcal/mol, also compares well
with the direct calculation of the reorganization energy from MD
simulations, where the value of 60.9 kcal/mol was reported
\cite{Ungar:99}. The electrostatic forces in those simulations were
cut off at distances greater than 10.1 \AA. The cutoff is expected to
lower the reorganization energy compared to that of an infinite
system.  In order to estimate the effect of the interaction cutoff, we
have calculated the reorganization energy for a fictitious solute with
the distance 10.1 \AA{} added to the radius of each atom exposed to the
solvent. This contribution amounts to 7.1 kcal/mol. Column 6 in Table
\ref{tab:1} shows the results of calculations when the $k$-dependent
polarization structure factors are replaced by their $k=0$ values. The
gap in $\lambda_s$ values between columns 5 and 6 thus quantifies the
contribution of the non-local part of solvent response to the
reorganization energy. The last (10) column in Table \ref{tab:1} shows
the PPSF calculations using parameters of ambient water.  In these
calculations, the gas phase dipole moment $m=1.87$ D is renormalized
by the polarizability effect to give $m'=2.43$ D (Wertheim's 1-RPT
formalism \cite{Wertheim:79,DMjpca:04}).  Despite this
renormalization, $\lambda_s$ in this calculation is substantially ($\simeq$ 30 \%)
smaller than in the calculations using parameters of TIP3P water.
TIP3P water thus appears to produce stronger solvation than ambient
water.

\begin{table}
\caption{{\label{tab:2}}Reorganization energy (kcal/mol) and 
          reorganization entropy (e.u., cal K$^{-1}$mol$^{-1}$) at T=298 K 
          of complex \textbf{1}
         (experimental parameters for ambient water) calculated with the PPSF
         for the structure factors.}
\begin{tabular}{cccccccc}
\hline
\hline
$\epsilon_{\infty}$ & $\epsilon_s$ & $\lambda_s$\footnotemark[1] & $S_{\lambda}$\footnotemark[1] &  $\lambda_s$\footnotemark[2]
        & $S_{\lambda}$\footnotemark[2] & $\lambda_s$\footnotemark[3] & $\lambda_s$\footnotemark[4]\\
\colrule
1.0 & 78.0 & 52.97 & 46.50 & 39.49 & $-8.14$ & 81.12 & 43.64\\
1.2 & 78.0 & 48.84 & 52.17 & 32.84 & $-4.85$ & 69.61 & 37.44\\
1.4 & 78.0 & 46.41 & 61.52 & 28.11 & $-2.88$ & 61.39 & 33.01\\
1.6 & 78.0 & 45.32 & 70.66 & 24.52 & $-1.60$ & 55.23 & 29.69\\
1.8\footnotemark[5]& 78.00 & 45.15 & 80.01 & 21.74 & $-0.71$ & 50.44 &27.11 \\
2.0 & 78.0 & 45.68 & 90.12 & 19.52 & $-0.15$ & 46.60 & 26.05\\
\hline
1.0\footnotemark[6] & 97.5\footnotemark[7] & 64.11\footnotemark[8] & 
     84.13\footnotemark[9]  & 39.56\footnotemark[2]  & 2.96\footnotemark[2] & 81.44 & 43.79\\   
\hline
\hline
\end{tabular}
\footnotetext[1]{NRFT with the PPSF for ambient water with varying polarizability $\alpha$.}
\footnotetext[2]{Calculated with $S^L(k)=S^L(0)$ and $S^T(k) =S^T(0)$.}
\footnotetext[3]{DelPhi calculation with the vdW cavity. }
\footnotetext[4]{DelPhi calculation with the solvent-accessible cavity. }
\footnotetext[5]{Parameters corresponding to water at ambient conditions.}
\footnotetext[6]{TIP3P water.}
\footnotetext[7]{From present MD simulations. This value is in good agreement with
                 $\epsilon_s=97.0$ reported in the literature \cite{Guillot:02}. } 
\footnotetext[8]{Calculated for TIP3P water using the PPSF.}
\footnotetext[9]{$d\epsilon_s/dT=-0.654$ K$^{-1}$ from MD simulations is used; this value turns 
                 to be higher than experimental $d\epsilon_s/dT=-0.398$ K$^{-1}$. } 
\end{table}

Table \ref{tab:1} also presents two components of the solvent
reorganization energy produced by solvent quadrupoles: $\lambda_q$ is the
second cumulant of the coupling of the solute electric field gradient
to solvent quadrupole moment \cite{DMjcp2:99,DMjcp4:05} whereas
$\lambda_{pq}$ is a mixed term arising from correlated fluctuations of
dipoles and quadrupoles positioned at \textit{different} solvent
molecules \cite{DMjcp2:99,DMjcp:05QP}. The resulting solvent
reorganization energy is the sum of the dipolar component 
$\lambda_p$ and two quadrupolar components:
\begin{equation}
  \label{eq:lambda}
  \lambda_s = \lambda_p + \lambda_{pq} + \lambda_q .
\end{equation}

The problem of quadrupolar solvent reorganization has recently
attracted much attention \cite{Perng:96,Perng2:96,DMjcp2:99,Jeon:01}
in connection with new experimental data showing appreciable solvent
reorganization in non-dipolar solvents
\cite{Britt:95,Reynolds:96,Kulinowski:95,Khajehpour:00,Read:00}.
However, the components $\lambda_q$ and $\lambda_{pq}$ constitute only a small
fraction of the overall reorganization energy despite a relatively
high reduced quadrupole of water, $\beta Q^2 / \sigma^5 = 1.1$ (cf.\ to
$(Q^*)^2= 0.13$ of acetonitrile).  For the rest of the paper we will
therefore assume
\begin{equation}
  \label{eq:app}
  \lambda_s \simeq \lambda_p .
\end{equation}
We note that the value $\lambda_s=69.7$ kcal/mol calculated for TIP3P water
with the account of water quadrupoles is in remarkable agreement with
$\lambda_s=68$ kcal/mol obtained by correcting the simulated values
\cite{Ungar:99} by the finite-size cutoff effects.

The dependence of $\lambda_s$ and the reorganization entropy on $\epsilon_{\infty}$ are
given in Table \ref{tab:2}.  In these calculations, the vacuum dipole
moment of water, 1.83 D, was held constant along with the total
dielectric constant $\epsilon_s=78.0$. The change in $\epsilon_{\infty}$ was achieved by
varying the polarizability $\alpha$ according to the Clausius-Mossotti
equation
\begin{equation}
  \label{eq:4-2}
  \frac{\epsilon_{\infty}-1}{\epsilon_{\infty}+2} = 8\eta\alpha/ \sigma^3  , 
\end{equation}
where $\eta=(\pi /6)\rho\sigma^3$ is the solvent packing fraction.  

Two drastically different predictions for the effect of solvent
polarizability on $\lambda_s$ can be found in the literature. The classical
Marcus two-sphere model \cite{Marcus:93} predicts a drop of $\lambda_s$ by
about a factor of 0.6 when going from $\epsilon_{\infty}=1.0$ to $\epsilon_{\infty}=1.8$. On
the other hand, simulations using non-polarizable and polarizable
versions of the water force field predict almost no dependence of
$\lambda_s$ on solvent polarizability \cite{Bader:96,comBaderBerne}. The actual situation
is in between of the two extremes. The reorganization energy does drop
with increasing $\epsilon_{\infty}$, but not as much as is predicted by continuum
models \cite{DMjpca:04}. On the other hand, the change is sufficient to make
simulations based on non-polarizable solvent models unreliable.

\begin{figure}[htbp]
  \centering \includegraphics*[width=6cm]{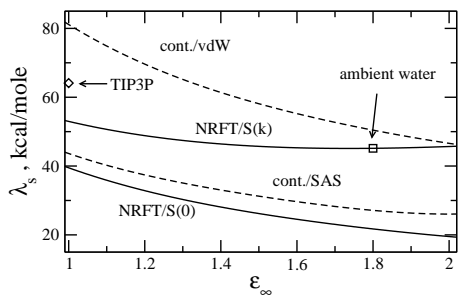}
  \caption{$\lambda_p$ vs $\epsilon_{\infty}$ calculated for complex \textbf{1} by using the nonlocal polarization 
    response theory (NRFT, solid lines). The dashed lines refer to the
    numerical solution of the Poisson equation with the vdW
    (cont./vdW) and SAS (cont./SAS) cavities. The diamond and square
    indicate TIP3P and ambient water, respectively.}
  \label{fig:11poly}
\end{figure}

The situation for the dependence of $\lambda_s$ on $\epsilon_{\infty}$ is illustrated in
Fig.\ \ref{fig:11poly}, where continuum results for complex \textbf{1}
obtained with the DelPhi Poisson-Boltzmann solver \cite{delphi:02} are
compared to the calculations within the NRFT. The dielectric
calculations with the vdW dielectric cavity (denoted ``cont./vdW'' in
Fig.\ \ref{fig:11poly}) show a substantial drop of $\lambda_s$ with $\epsilon_{\infty}$.
The dependence on $\epsilon_{\infty}$ is much weaker in the NRFT (see also Table
\ref{tab:2}). The weak dependence of $\lambda_s$ on $\epsilon_{\infty}$ is the result of
the cancellation of two competing factors: the decrease of the
longitudinal structure factor in the range of small $k$-values with
increasing $\epsilon_{\infty}$ (Fig.\ \ref{fig:4poly}) compensated by an increase
in $y_{p}$ due to higher solvent dipole $m'$ in more polarizable
solvents. We note that this cancellation is strongly affected by the
$k$-dependence of the polarization structure factors in the range of
small $k$-values contributing to the $k$-integral and cannot be
reduced to the cancellation of the $y_p$ factor in $\lambda_s$ [Eqs.\ 
(\ref{eq:2-10}) and (\ref{eq:2-12})] with $y_p$ in the denominator in
Eq.\ (\ref{eq:3-11}), resulting in the Pekar factor of continuum
electrostatics.

The continuum limit of the NRFT is obtained when the dependence on the
wavevector $k$ is neglected in the solvent structure factors and one
assumes $S^{L,T}(k)\simeq S^{L,T}(0)$ and $S^{L,T}_n(k)\simeq S^{L,T}_n(0)$.
When this assumption is incorporated in the microscopic calculations
(marked NRFT/$S(0)$ in Fig.\ \ref{fig:11poly}), the resultant
reorganization energy gains the strong dependence on $\epsilon_{\infty}$
characteristic of continuum theories. The continuum limit of the
microscopic theory corresponds, however, to the dielectric cavity
coinciding with the SAS. The corresponding DelPhi calculation (marked
cont./SAS in Fig.\ \ref{fig:11poly}) indeed goes parallel with the
continuum limit of the NRFT.  The distinction between these two
results arises from the mean-field approximation used in the NRFT
formulation and different handling of the polarizability effects in
the two formulations (additive in the continuum and non-additive in
the microscopic formulation \cite{DMjcp:05unpubl}). Note that the
mean-field approximation is more accurate, when compared to the exact
solution of the Li-Kardar-Chandler equation, in the full
microscopic formulation than in its continuum limit \cite{DMjcp2:04}.
The exact formulation of the theory, which does not involve the
mean-field approximation, gives the solution of the Poisson equation in
its continuum limit.

The numerical values for the reorganization energies shown in Fig.\ 
\ref{fig:11poly} are given in Table \ref{tab:2}. The comparison
between the microscopic and continuum calculations is instructive. At
$\epsilon_{\infty}=1$, $\lambda_s$ from the vdW continuum is much higher than the
microscopic calculation, while $\lambda_s$ from the SAS continuum is close
to the microscopic result.  With increasing $\epsilon_{\infty}$, on the other
hand, $\lambda_s$ from the vdW continuum falls down almost to the
microscopic value. The continuum calculation with the vdW cavity may
thus appear in a reasonable accord with microscopic calculations or
experiment due to the mutual cancellation of errors.

Along with reorganization energies, Table \ref{tab:2} lists
reorganization entropies $S_{\lambda}$ [Eq.\ (\ref{eq:4-1})].  Note that
$S_{\lambda}$ obtained from the PPSF calibrated on TIP3P water is in a
reasonable agreement with the corresponding value obtained with the
structure factors from MD simulations: 84.1 e.u.\ and 69.9 e.u.,
respectively.  The dielectric continuum calculation gives the wrong
sign for the entropy in accord with previous reports
\cite{DMcp:93,DMjpcb:99}. Also the magnitude of $S_{\lambda}$ is
substantially higher in the microscopic theory than in the continuum
calculation (cf.\ columns 4 and 6 in Table \ref{tab:2}).

A similar trend is seen for the reaction free energy gap (Table
\ref{tab:3}) for which the reaction entropy is defined as
\begin{equation}
  \label{eq:4-3}
  \Delta S_s = - \left(\partial \Delta G_s / \partial T \right)_P .
\end{equation}
Although the sign of $\Delta S_s$ is correct in the continuum calculations,
the entropy magnitude is much lower than in the NRFT, similar to a
previous report for a different ET system \cite{DMjpcb:99}, where $\Delta S_s$ was experimentally
obtained from temperature dependent absorption and emission
charge-transfer bands.  Since the analytical theory seems to be
consistent with the computer experiment, one needs a test against
experimental data.  Unfortunately, experimental evidence on the
solvent entropic effects on ET reactions is very limited (see Ref.\ 
\onlinecite{ZimmtWaldeck:03} for a recent review).

\begin{table}
\caption{{\label{tab:3}}Solvation Gibbs energy (kcal/mol) and solvation entropy 
                        (e.u., cal K$^{-1}$mol$^{-1}$).}
\begin{tabular}{ccccccc}
\hline
\hline
$\epsilon_{\infty}$ & $G_{s,1}$\footnotemark[1] & $S_{s,1}$\footnotemark[1] 
     &  $G_{s,2}$\footnotemark[2]
        & $S_{s,2}$\footnotemark[2] & $\Delta G_s$\footnotemark[3] 
        & $\Delta S_s$\footnotemark[3]\\
\colrule
1.0 & $-236.01$ & $-198.30$ & $-183.36$               & $-132.68$ & 52.65 & 65.62 \\
1.2 & $-245.13$ & $-228.90$ & $-189.28$                & $-151.94$ & 55.85 & 76.96 \\
1.4 & $-255.10$ & $-262.71$ & $-195.75$                & $-173.15$ & 59.35 & 89.56 \\
1.6 & $-266.00$ & $-299.28$&  $-202.79$                 & $-195.60$ & 63.21 & 103.68 \\
1.8\footnotemark[4] & $-277.88$ & $-337.52$ & $-210.43$& $-219.15$ & 67.45 & 118.37\\
2.0 & $-290.77$ &  $-378.20$ & $-218.66$               & $-243.41$ & 72.11 & 134.79 \\
\hline
1.8\footnotemark[5] & $-325.1$  &$-2.9 $  & $-242.1$ & $-2.0 $ & 83.0 &  1.1  \\
1.8\footnotemark[6] & $-196.0$   &$-1.46$  & $ -277.88$ &  $-1.21$  & 67.45  &  0.26 \\
\hline
\hline
\end{tabular}\\
\footnotetext[1]{Gibbs energy and solvation entropy in the initial ET state.}
\footnotetext[2]{Gibbs energy and solvation entropy in the final ET state.}
\footnotetext[3]{$\Delta G_s= G_{s,2} - G_{s,1}$, $\Delta S_s = S_{s,2} - S_{s,1}$.}
\footnotetext[4]{Parameters corresponding to water at ambient conditions.}
\footnotetext[5]{Continuum calculation (DelPhi \cite{delphi}) with solute's 
                 vdW cavity}  
\footnotetext[6]{Continuum calculation (DelPhi \cite{delphi}) with solute's
                 SA cavity}
\end{table}

\begin{figure}[tbh]
    \includegraphics*[width=6cm]{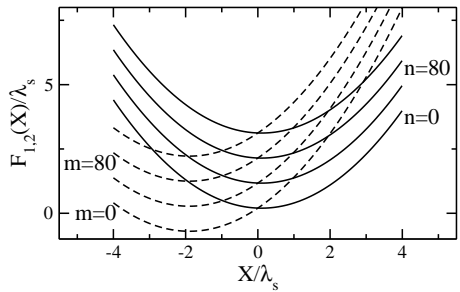}
    \caption{Diabatic initial (solid lines) and final (dashed lines)
      curves obtained from Eq.\ (\ref{eq:1-20}) for parameters of DSA
      \textbf{1}. Curves marked with $n$ and $m$ indexes refer to
      vibrational states of the initial and final states,
      respectively.  Relative energies are drawn to scale, based on 
      $\lambda_s$=45.15 kcal/mol, $\Delta G = - 41.2$ kcal/mol, $\hbar \omega_v$=400
      cm$^{-1}$.}
    \label{fig:12poly}
\end{figure}

\subsection{ET rate constant}
\label{sec:5-2}
The calculations of the temperature dependent reorganization energy
and equilibrium energy gap can be compared to experimental Arrhenius
law measurements \cite{Ogawa:93} for complex \textbf{1}.  Transition
metal-based charge-transfer complexes are commonly characterized by
metal-ligand vibrational frequencies \cite{Ungar:99} in the range
$\omega_v\simeq300-500$ cm$^{-1}$, substantially lower than frequencies
$\omega_v\simeq1100-1500$ cm$^{-1}$ normally assigned to C$-$C skeletal
vibrations of organic donor-acceptor complexes. Therefore, Eqs.\ 
(\ref{eq:0}), (\ref{eq:0-1}), and (\ref{eq:1-20}) with the full
quantum-mechanical description of vibrations and temperature-induced
populations of vibrational states should be used for the ET rate in
complex \textbf{1}. Unfortunately, our calculations provide only the
solvent component of the free energy gap. Its gas-phase component is
unknown and the electronic coupling entering the Golden Rule ET rate
in Eq.\ (\ref{eq:0}) is known with uncertainty \cite{Ungar:99}.  These
two parameters ($\Delta G_g$ and $V_{12}$) were varied in fitting the
experimental activation enthalpy $\Delta H^{\dag}=9.5$ kcal/mol and the
experimental activation entropy $\Delta S^{\dag}/k_{\text{B}}=-5.6$ e.u.\ 
\cite{Ogawa:93}.  Note that the experimental quantity is an effective
entropy, including contributions due to the electronic coupling
element as well as solvation and inner-sphere vibrational modes
\cite{Ungar:99}. The Arrhenius analysis was performed by the linear
regression of $\ln(k_{ET}/T)$ vs $1/T$ based on the transition-state
expression
\begin{equation}
  \label{eq:5-1}
  k_{\text{ET}} = \frac{k_{\text{B}}T}{h} e^{-\Delta G^{\dag}(T)/k_{\text{B}}T }.
\end{equation}

The rate constant was calculated based on Eqs.\ (\ref{eq:0}) and
(\ref{eq:1-20}), with $\lambda_s$ and $\Delta G_s$ varied linearly with
temperature using the calculated entropies (Tables \ref{tab:2} and
\ref{tab:3}). The results of calculations are listed in Table
\ref{tab:4}. The fitted electronic coupling $V_{12}$ falls in the
range of values given by electronic structure calculations
\cite{Ungar:99} using the semiempirical INDO/s model by Zerner and
co-workers \cite{Zerner:80}. The equilibrium gap obtained from the fit
is appreciably more negative than $\Delta G \simeq -25.4$ kcal/mol
estimated from the redox potentials of separate donor and acceptor
sites, based on the high spin ground state of the Co$^{2+}$ product
(it has been argued \cite{Ungar:99} that the less exothermic low spin
Co$^{2+}$ product may be the relevant one in the experimentally
observed process). Neglecting the vibrational excitations in the
analysis (0-0 transition only) results in a much lower activation
enthalpy and a substantially more negative activation entropy (second
row in Table \ref{tab:4}).

\begin{figure}[htbp]
  \centering
  \includegraphics*[width=6cm]{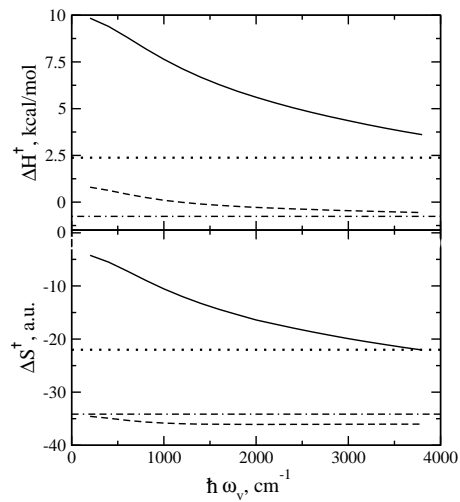}
  \caption{Enthalpy and entropy of activation of DSA complex \textbf{1} 
    vs $\hbar\omega_v$. The solid lines indicate the calculations according to
    Eq.\ (\ref{eq:1-20}) with $\lambda_s$ and $\Delta G_s$ varying linearly with
    temperature based on the corresponding entropies from Tables
    \ref{tab:2} and \ref{tab:3}.  The dashed line indicate the same
    calculation with $\lambda_s$ and $\Delta G_s$ fixed at their 298 K values.
    The dash-dotted ($S_{\lambda} = 0$, $\Delta S_s = 0$) and dotted ($S_{\lambda}\neq 0$,
    $\Delta S_s\neq 0$) lines for the activation enthalpy and entropy refer to
    the $\omega_v\to \infty$ limit corresponding to Eq.\ (\ref{eq:1-12}) with
    $\lambda_v=0$ (no vibrational excitations). }
  \label{fig:13poly}
\end{figure}

The relatively low frequency of metal-ligand vibrations in transition
metal complexes results in a dense manifold of vibrational levels
(Fig.\ \ref{fig:12poly}) which are partially populated at room
temperature.  The change of the vibrational populations with
temperature may result in a contribution to the overall activation
entropy \cite{Brunschwig:80}. This, however, does not happen for
complex \textbf{1} when $\lambda_s$ and $\Delta G_s$ are fixed at their 298 K
values. The dashed lines in Fig.\ \ref{fig:13poly} show the enthalpy
and entropy of activation as a function of the vibrational frequency
at constant temperature and $\lambda_v$. Increasing the vibrational
frequency makes vibrational excitations less accessible, but this is
seen to have little effect on the activation entropy and enthalpy.

This situation changes when the temperature dependence of $\lambda_s$ and $\Delta
G_s$ is included in the calculations of the Arrhenius activation
parameters.  In this case, the temperature dependence of the ET energy
gap results in a change of the vibrational quantum numbers 
corresponding to the maximum vibronic contribution. The splitting of the
activation barrier into the entropic and enthalpic contribution then
becomes sensitive to the choice of $\omega_v$ (Fig.\ \ref{fig:13poly}, solid
lines).  This sensitivity may be important for the interpretation of
experimental data since the correct definition of the effective
vibrational frequency [Eq.\ (\ref{eq:1-14})] increases in importance once
the temperature dependence of the solvation parameters is introduced
into the analysis of reaction rates.  The classical Marcus-Hush
equation with $\lambda_v=0$ replaces the sum over all possible vibronic
transitions with a single 0-0 transition. The result is a
significantly lower enthalpy and more negative entropy of activation
(Table \ref{tab:4}).

\begin{table}[htbp]
  \centering
  \caption{{\label{tab:4}}Parameters for complex \textbf{1} at T=298 K.}
  \begin{tabular}{ccccccc}
  \hline
Level &  $V_{12}$   & $\Delta G$    & $\hbar\omega_v$    & $\lambda_v$     & $\Delta S^{\dag}$ & $\Delta H^{\dag}$   \\
     & cm$^{-1}$ & kcal/mol & cm$^{-1}$  & kcal/mol &  e.u.   &  kcal/mol \\
  \hline
Full &  0.07\footnotemark[1]  & $-41.2$\footnotemark[1] & 400 & 16.1\footnotemark[2] &
  $-5.5$\footnotemark[3]  & 9.4\footnotemark[3]  \\
$\lambda_v=0$\footnotemark[4]  & 0.07  &  $-25.4$\footnotemark[5]  & --  & 0.0  
                        & $-25.0$\footnotemark[4]  & 2.0\footnotemark[4] \\  
  \hline
  \end{tabular}
  \footnotetext[1]{Obtained from fitting the experimental Arrhenius dependence. $S_{\lambda}$ and $\Delta S_s$ are calculated within the PPSF with parameters corresponding to ambient water.}
  \footnotetext[2]{From Ref.\ \onlinecite{Ungar:99}.}
  \footnotetext[3]{Experimental values from Ref.\ \onlinecite{Ogawa:93}.}
  \footnotetext[4]{Obtained by neglecting intramolecular vibrations ($\lambda_v=0$) in
                   Eq.\ (\ref{eq:1-20}). In this limiting case, the $V_{12}$ value was taken
                   from the value obtained from the full analysis. }
  \footnotetext[5]{Estimated from redox potentials of separate donor and acceptor, 
                   Ref.\  \onlinecite{Ogawa:93}.}
  
\end{table}

\begin{figure}[htbp]
  \centering
  \includegraphics*[width=6cm]{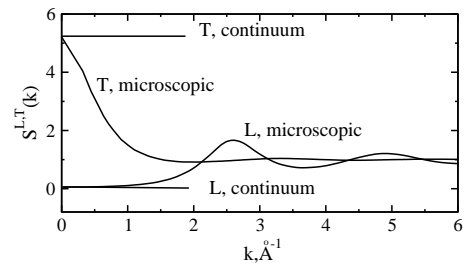}
  \caption{Microscopic structure factors and their continuum limits. }
  \label{fig:13}
\end{figure}

\section{Discussion}
\label{sec:6}
The most relevant question in comparing microscopic solvation theories
with the dielectric continuum approximation is why the latter has
allowed to describe so many systems after proper parameterization of
dielectric cavities, despite drastic approximations involved.  The
microscopic NRFT formulation contains dielectric continuum as its
limit, allowing us to address this question. The continuum limit is
obtained by neglecting the spatial correlations between solvent
dipoles, i.e.\ by neglecting the $k$-dependence in the polarization
response functions.  This implies that polarization structure factors
are replaced by their $k=0$ values (Fig.\ \ref{fig:13}). This
replacement is not a good approximation for the transverse structure
factor, which changes quite sharply even for small $k$-values, but may
be a reasonable approximation for the longitudinal structure factor,
which is relatively flat in the range of $k$-values significant for
solvation thermodynamics. However, for most charge configurations,
even for the point dipole \cite{DMjcp1:04}, the contribution of
transverse polarization to the solvation free energy is relatively
small \cite{DMjcp2:04} ($\simeq 10$\% in our calculations for complex
\textbf{1} in water). Therefore, the inaccurate continuum
approximation for the transverse structure factor does not
significantly affect the results of calculations.

The continuum estimates for the polarization structure factors result
in the following inequalities between the continuum and microscopic
longitudinal and transverse components of the reorganization energy
\begin{equation}
  \label{eq:6-1}
  \lambda_s^{L,\text{cont}} < \lambda_s^L,\quad \lambda_s^{T,\text{cont}} > \lambda_s^T . 
\end{equation}
The sharp change of the transverse structure factor at small
$k$-values is responsible for a substantial overestimate of the
transverse component of solvation by continuum models
\cite{DMjcp1:04,DMjcp2:04}. This overestimate manifests itself in
solvation dynamics. The transverse polarization dynamics is much
slower than the longitudinal polarization dynamics \cite{Bagchi:91}.
Therefore, continuum models predict biphasic solvation dynamics with
an appreciable slow component due to transverse polarization
relaxation.  This slow component is not observed in computer
simulations of solvation dynamics \cite{Kumar:95} and it does not show
up in the microscopic calculations reported in Ref.\ \onlinecite{DMjcp1:05}.

\begin{figure}[htbp]
  \centering \includegraphics*[width=6cm]{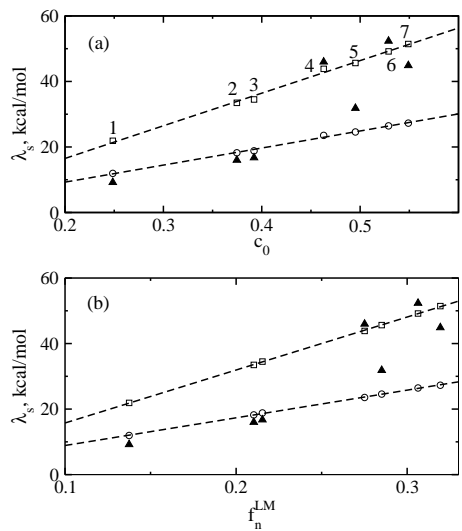}
   \caption{Reorganization energies in polar solvents vs the Pekar factor [Eq.\ (\ref{eqPekar})]
     (a) and the Lippert-Mataga polarity parameter [Eq.\ (\ref{eq:1})]
     (b). The open points indicate the DelPhi calculations with vdW
     (squares) and SAS (circles) molecular surfaces. The closed points
     (triangles) refer to calculations with the NRFT.  Numbers on the
     plot indicate: chloroform (1), tetrahydrofuran (2), methylacetate
     (3), N,N-dimethylformamide (4), acetone (5), acetonitrile (6),
     water (7). }
  \label{fig:14}
\end{figure}

The relatively flat form of the longitudinal structure factors at low
$k$-values does not mean that replacing $S^{L,T}(k)$ by $S^{L,T}(0)$
gives accurate numbers for the solvation free energy and/or the
reorganization energy. A moderate increase of $S^L(k)$ in the range of
wavevectors contributing to the $k$-integral substantially affects the
calculated values of solvation free energies (cf.\ columns 5 and 6 in
Table \ref{tab:1}). Moreover, the gap between the microscopic and
continuum values changes with the solvent dielectric parameters (see,
e.g., Fig.\ \ref{fig:11poly}). This observation practically means that
there is fundamentally no unique scheme for defining the dielectric
cavity applicable to all solvent polarities.

The dominance of longitudinal polarization fluctuations in solvation
thermodynamics is also responsible for experimentally observed linear
trends of the reorganization energy with the Pekar factor
\cite{Powers:80,Grampp:84,Hupp:93} [Eq.\ (\ref{eqPekar})]. Even at the
continuum level, the polarization response function for a solute of
complex shape is not represented by the Pekar factor appearing in the
longitudinal projection of the solvent response function \cite{BES:86}.
However, large separation of charges is responsible for the
predominantly longitudinal response of the solvent, and continuum
reorganization energies calculated for complex \textbf{1} in polar
solvents correlate well with the Pekar factor (Fig.\ \ref{fig:14}a).
If fact, an equally good correlation is seen in respect to the
Lippert-Mataga polarity parameter commonly used for solvation of
dipoles (Fig.\ \ref{fig:14}b):
\begin{equation}
  \label{eq:1}
   f_n^{\text{LM}} = \frac{\epsilon_s -1}{2\epsilon_s +1} - \frac{\epsilon_{\infty}-1}{2\epsilon_{\infty}+1} . 
\end{equation}
The use of a particular parameter does not therefore tell much about
the nature of the solute charge distribution and, obviously, reflects
a linear relation between $c_0$ and $f_n^{\text{LM}}$ for common
solvents.

\begin{figure}[htbp]
  \centering \includegraphics*[width=6cm]{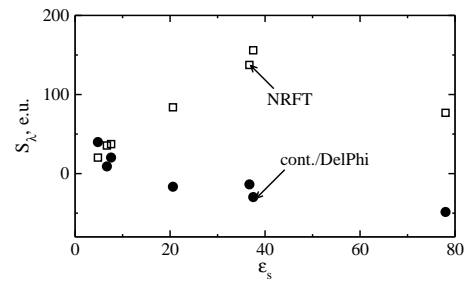}
  \caption{Reorganization entropies from continuum calculations (DelPhi \cite{delphi} with 
    the vdW surface, closed circles) 
    and by the NRFT (open squares). Points refer to the same solvents
    as in Fig.\ \ref{fig:14}. }
  \label{fig:15}
\end{figure}

The results of the current microscopic calculations are shown by
triangles in Fig.\ \ref{fig:14}. These numbers do not exhibit a
linear dependence, although the extent of scatter is not uncommon for
ET experiment. The comparison of the continuum and microscopic
dependence on the solvent polarity does not permit a clear
distinction between the two formulations. Where the distinction becomes
clear is for the reorganization entropy in strongly polar solvents.
Figure \ref{fig:15} shows reorganization entropies $S_{\lambda}$ calculated
in continuum (DelPhi \cite{delphi} Poisson-Bolzmann solver) and
microscopic (NRFT) theories. The continuum calculation reflects the
temperature variation of the Pekar factor $c_0$:  
\begin{equation}
- (\partial c_0/ \partial T)_P =  \epsilon_{\infty}^{-2} (\partial \epsilon_{\infty}/ \partial T)_P - \epsilon_s^{-2} (\partial \epsilon_s/ \partial T)_P
\end{equation}
In low-polarity solvents, $c_0$ is mostly influenced by the static
dielectric constant, which has a negative temperature derivative. The
continuum reorganization entropy (closed circles in Fig.\ 
\ref{fig:15}) is positive and is close to the microscopic result
(open squares in Fig.\ \ref{fig:15}).  The continuum estimate of the
temperature variation of $\lambda_s$ in low-polarity solvents thus gives a
semi-quantitative account of the experimental observations
\cite{Liang:89}. In strongly polar solvents, the temperature
derivative of $c_0$ is mostly influenced by the high-frequency
dielectric constant, and continuum $S_{\lambda}$ is nagative.  In this case,
the predictions of the continuum model significantly depart from both
the microscopic calculations and many experimental measurements
\cite{Grampp:84,Derr:98,Nelsen:99,Derr:99,DMjpcb:99,Vath:00,Zhao:01},
showing positive reorganization entropies.

The microscopic calculations presented here show a relatively weak
dependence of the reorganization energy on the solvent high-frequency
dielectric constant, in qualitative accord with available computer
simulation data \cite{Bader:96,Ando:01,DMjpca:04}. Testing this
theoretical prediction experimentally may become problematic because
of the narrow range of $\epsilon_{\infty}$ values available for common polar
solvents.  We note, however, that the problem of the weak dependence
of the reorganization energy on $\epsilon_{\infty}$ is related to the problem of
correct sign of the reorganization entropy.  The strong dependence of
the continuum reorganization energy on $\epsilon_{\infty}$ is one of major
factors shifting the continuum reorganization entropy to the range of
positive values.

The calculations of the quadrupolar component of the solvent
reorganization energy presented here confirm the conclusion previously
reached for Stokes shifts in coumarin-153 optical dye
\cite{DMjpca:01}: quadrupolar solvation is insignificant in most
commonly used polar solvents, and the dipolar approximation for the
solvent charge distribution is sufficient for most practical
calculations.

\acknowledgments D.V.M.\ thanks the Donors of The Petroleum Research
Fund, administered by the American Chemical Society (39539-AC6), for
support of this research.  M.D.N.\ was supported by DE-AC02-98CH10886
at Brookhaven National Laboratory.  The authors are grateful to Prof.\ 
G.\ A.\ Voth for sharing the structural data on the polypeptide-linked
donor-acceptor complex. This is publication \#596
from the ASU Photosynthesis Center.

\appendix

\section{Simulation and analysis.}
\label{A1}
The MC simulations of dipolar-polarizable hard sphere solvents shown
in Fig.\ \ref{fig:8poly} were done as described in Ref.\ 
\onlinecite{DMjpca:04}.  Simulations of $6\times10^{5}$ cycles long were
run for 1372 polarizable molecules with periodic boundary conditions
and the reaction field cutoff of dipole-dipole interactions.  The MD
simulations were carried out with the force field of 3-site
acetonitrile (ACN3) by Edwards, Madden, and
McDonald \cite{acnMadden:84} and the 3-site model of water (TIP3P) by
Jorgensen \textit{et al}. \cite{jorg:83} (Table \ref{tab:22}).  The
site-site interaction potential is given by the sum of the
Lennard-Jones (LJ) and Coulomb interaction potentials:
\begin{equation}
\label{eq:a3-1}
E_{\alpha\beta}=4\varepsilon_{\alpha\beta}\left[\left(\dfrac{\sigma_{\alpha\beta}}{r_{\alpha\beta}}\right)^{12}-
       \left(\dfrac{\sigma_{\alpha\beta}}{r_{\alpha\beta}}\right)^6\right]+\dfrac{q_{\alpha}q_{\beta}}{r_{\alpha\beta}},
\end{equation} 
where the LJ parameters are taken according to the Lorentz-Bertholet
rules: $\varepsilon_{\alpha\beta}=\sqrt{\varepsilon_{\alpha}\varepsilon_{\beta}}$ and $\sigma_{\alpha\beta}=(\sigma_{\alpha}+\sigma_{\beta})/2$.  All
simulations were done with the DL\_POLY molecular dynamics
package \cite{dlpoly2:96}. We run two sets of MD simulations in the
temperature range from 288 K to 308 K with a 5 K step. The timestep in
each simulation is 5 fs.  All MD simulation are 20 ns long.

 \begin{table}[tbh]
\caption{{\label{tab:5}}Force field parameters for use in the simulation.}
\begin{tabular}{cccc}
\hline
Atomic interaction site &$\sigma_{\alpha} $/\AA&$\varepsilon_{\alpha}\times 10^3$/ (kcal/mol)  & $q_{\alpha}$/ e  \\
\hline 
TIP3P water\footnotemark[1] &  &  &    \\ 
 O&3.15 &152.10& $-0.834$   \\ 
Acetonitrile\footnotemark[2] &  &  &    \\ 
 Me&  3.6& 379.55&   0.269 \\ 
 C&  3.4&  99.36&    0.129\\ 
 N &  3.3& 99.36 &   $ -0.398$\\
 \hline 
 \footnotetext[1]{ r$_{\mathbf{OH}}$=0.9572 \AA, $\angle$ HOH=104.52$^\circ$}
 \footnotetext[2]{r$_{\mathbf{MeC}}$=1.46 \AA, r$_{\mathbf{CN}}$=1.17 \AA}
 \end{tabular} 
 \label{tab:22}
\end{table}

We used the Nos\'e-Hoover thermostat \cite{hoover:85} for the ACN3
simulations with the relaxation parameter of 0.5 fs. This value
ensures good stabilization of the total system energy. The energy
drift for ACN3 is only about 0.1\%. The simulation box was constructed
to include 256 ACN3 molecules in a cube with the side length
$L=28.2025$ \AA{} at T=298 K to reproduce the experimental mass density of
acetonitrile, $\rho_M$=0.777 g/cm$^3$. The side length is adjusted at
each temperature to account for temperature expansion with the
experimental volume expansion coefficient $\alpha_p=1.38\times10^{-3}$ K$^{-1}$.
        
In simulations of TIP3P water, 256 molecules reside in a cube with the
side length of $L=19.7744$ \AA{} at 298 K. The system is coupled to the
Berendsen \cite{berend:84} thermostat with the relaxation time of 0.1
fs. The drift in total energy of about 0.1 \% is observed.  The liquid mass density
$\rho_M=0.9896$ g/cm$^3$ and the volume expansion coefficient $\alpha_p=2.96\times
10^{-3}$ K$^{-1}$ are taken from Ref.\ \onlinecite{xxx:20nsSIM}.  The
latter value is close to the experimental expansion coefficient of
ambient water, $\alpha_p= 2.6\times10^{-3}$ K$^{-1}$.

The cutoff for short-range LJ interaction is 13 \AA{} for ACN3 and 9
\AA{} for TIP3P. For long-range Coulomb interactions, Ewald
summation from DL\_POLY \cite{Allen:96} is used for ACN3 and smoothed
particle mesh (SPME) \cite{spme:95} Ewald is adopted for TIP3P.  Ewald summation
parameters are the convergence parameter $\alpha$ and the
maximum wavenumber $k^{max}_{x,y,z}$. The parameter sets $\alpha =0.24 $
\AA$^{-1}$, $k^{max}_{x,y,z}=7$ \AA$^{-1}$, and $\alpha$ =0.35
\AA$^{-1}$, ${k}^{max}_{x,y,z}=8$ \AA$^{-1}$ were used for ACN3 and TIP3P respectively.

The structure factors have been calculated as the variance of 
longitudinal and transverse projections of the $\mathbf{k}$-space 
solvent polarization
\begin{equation}
\label{eq:a3-2}
  \mathbf{M}(k)=(1/m)\sum_{i=1}^N  \mathbf{m}_ie^{-\mathit{i}\mathbf{k}\cdot\mathbf{r}_i},
\end{equation}   
where $\mathbf{m}_i=\sum_a q_a \mathbf{r}_i^a $ is a dipole moment of the
$i$th molecule and the sum runs over the $N$ molecules in the
simulation box.  The static dielectric constant is given in terms of
the $k=0$ variance as follows \cite{Neumann:86}
\begin{equation}
\label{a3-3}
        \varepsilon_s=1+3 y \langle \mathbf{M(0)}^2\rangle/N ,
\end{equation} 
where $y=(4\pi/9)\rho m^2/k_{\text{B}} T $.

\bibliography{/home/dmitry/p/bib/chem_abbr.bib,/home/dmitry/p/bib/et,/home/dmitry/p/bib/solvation,/home/dmitry/p/bib/bioet,/home/dmitry/p/bib/etnonlin,/home/dmitry/p/bib/liquids,/home/dmitry/p/bib/dm,/home/dmitry/p/bib/dynamics,/home/dmitry/p/bib/comments,/home/dmitry/p/bib/photosynth}

\begin{thebibliography}{133}
\bibitem[{com({\natexlab{b}})}]{comPekar}
\bibinfo{note}{Different partitioning schemes of the solvent polarization into
  the fast and slow components can be implemented for solvation thermodynamics
  calculations \cite{Pekar:63,Brady:85,Aguilar:01}. Although they all give
  identical results for the thermodynamics of solvation, they will lead to
  different predictions regarding the solvation dynamics. The Pekar
  partitioning results in slow relaxation of the component of induced dipole
  ($\mathbf{p}_j^R$ in Eq.\ (\ref{eq:10-1-1})) in equilibrium with the field of
  permanent dipoles in the liquid.}
\bibitem[{com({\natexlab{c}})}]{comQ}
\bibinfo{note}{The quadrupole moment is a tensor; the scalar quadrupole moment
  is defined as $Q^2=(2/3)\mathbf{Q}:\mathbf{Q}$ \cite{Gubbins:84}.}
\end{thebibliography}
\end{document}